\documentclass[fleqn,usenatbib]{mnras}

\usepackage{newtxtext,newtxmath}

\usepackage[T1]{fontenc}

\DeclareRobustCommand{\VAN}[3]{#2}
\let\VANthebibliography\thebibliography
\def\thebibliography{\DeclareRobustCommand{\VAN}[3]{##3}\VANthebibliography}

\usepackage{graphicx}	
\usepackage{amsmath}	
\usepackage{threeparttable}
\usepackage{booktabs}
\usepackage{url}

\newcommand{\Tzerob}[1][days]   {$1994.2853 _{ - 0.003 } ^ { + 0.0028 }$~#1} 
\newcommand{\Pb}[1][days]   {$11.62071 _{ - 0.000106 } ^ { + 9.6e-05 }$~#1} 
\newcommand{\esinb}[1][ ]   {$-0.02 _{ - 0.16 } ^ { + 0.15 }$~#1} 
\newcommand{\ecosb}[1][ ]   {$0.06 _{ - 0.2 } ^ { + 0.19 }$~#1} 
\newcommand{\bb}[1][ ]   {$0.2 _{ - 0.14 } ^ { + 0.17 }$~#1} 
 
\newcommand{\rrb}[1][ ]   {$0.01957 _{ - 0.00063 } ^ { + 0.00062 }$~#1} 
\newcommand{\kb}[1][${\rm m\,s^{-1}}$]   {$2.53 _{ - 0.41 } ^ { + 0.44 }$~#1} 
\newcommand{\mpb}[1][$M_{\oplus}$]   {$9.82 _{ - 1.61 } ^ { + 1.71 }$~#1} 
\newcommand{\rpb}[1][$R_{\oplus}$]   {$3.05 \pm 0.11 $~#1} 
 
\newcommand{\eb}[1][ ]   {$0.046 _{ - 0.033 } ^ { + 0.058 }$~#1}

\newcommand{\ib}[1][deg]   {$89.26 _{ - 0.58 } ^ { + 0.51 }$~#1} 
 
\newcommand{\ab}[1][AU]   {$0.1049 _{ - 0.003 } ^ { + 0.0029 }$~#1}

\newcommand{\insolationb}[1][${\rm F_{\oplus}}$]   {$265.4 _{ - 20.4 } ^ { + 22.1 }$~#1} 
 
\newcommand{\denstrb}[1][${\rm g\,cm^{-3}}$]   {$0.551 _{ - 0.04 } ^ { + 0.039 }$~#1}

\newcommand{\ttotb}[1][hours]   {$5.59 _{ - 0.14 } ^ { + 0.13 }$~#1}

\newcommand{\denpb}[1][${\rm g\,cm^{-3}}$]   {$1.9 _{ - 0.36 } ^ { + 0.41 }$~#1}

\newcommand{\Tzeroc}[1][days]   {$2002.7715 _{ - 0.0015 } ^ { + 0.0016 }$~#1} 
\newcommand{\Pc}[1][days]   {$19.502104 _{ - 8.5e-05 } ^ { + 7.4e-05 }$~#1} 
\newcommand{\esinc}[1][ ]   {$0.05 _{ - 0.27 } ^ { + 0.16 }$~#1} 
\newcommand{\ecosc}[1][ ]   {$0.09 _{ - 0.23 } ^ { + 0.19 }$~#1} 
\newcommand{\bc}[1][ ]   {$0.3 _{ - 0.21 } ^ { + 0.18 }$~#1} 
\newcommand{\rrc}[1][ ]   {$0.03804 _{ - 0.00087 } ^ { + 0.00103 }$~#1} 
\newcommand{\kc}[1][${\rm m\,s^{-1}}$]   {$2.11 _{ - 0.42 } ^ { + 0.42 }$~#1} 
\newcommand{\mpc}[1][$M_{\oplus}$]   {$9.67 _{ - 1.92 } ^ { + 1.97 }$~#1} 
\newcommand{\rpc}[1][$R_{\oplus}$]   {$5.94 _{ - 0.16 } ^ { + 0.18 }$~#1} 
 
\newcommand{\ec}[1][ ]   {$0.074 _{ - 0.05 } ^ { + 0.072 }$~#1}

\newcommand{\ic}[1][deg]   {$89.21 _{ - 0.4 } ^ { + 0.53 }$~#1} 
 
\newcommand{\ac}[1][AU]   {$0.1482 _{ - 0.0042 } ^ { + 0.0041 }$~#1}

\newcommand{\insolationc}[1][${\rm F_{\oplus}}$]   {$133.1 _{ - 10.2 } ^ { + 11.1 }$~#1}

\newcommand{\ttotc}[1][hours]   {$6.5 _{ - 0.11 } ^ { + 0.16 }$~#1}

\newcommand{\qone}[1][]   {$0.39 _{ - 0.15 } ^ { + 0.32 }$~#1} 
\newcommand{\qtwo}[1][]   {$0.55 _{ - 0.24 } ^ { + 0.27 }$~#1}

\newcommand{\HARPSN}[1][${\rm km\,s^{-1}}$]   {$0.00042 _{ - 0.00029 } ^ { + 0.0003 }$~#1} 
\newcommand{\jHARPSN}[1][${\rm m\,s^{-1}}$]   {$2.74 _{ - 0.23 } ^ { + 0.25 }$~#1} 
\newcommand{\jtr}[1][]   {$0.00024 _{ - 1.4e-05 } ^ { + 1.3e-05 }$~#1}

\newcommand\vsini{\ensuremath{v \, \sin \, i_\star}} 
   
\newcommand{\kms}{\ensuremath{\text{km}\,\text{s}^{-1}}} 
\newcommand{\ms}{\ensuremath{\text{m}\,\text{s}^{-1}}}

\newcommand{\MEarth}{\ensuremath{M_{\oplus}}}	
\newcommand{\REarth}{\ensuremath{R_{\oplus}}}	

\newcommand{\tess}{\textit{TESS}}
\newcommand{\target}{HD\,152843}
\newcommand{\planetb}{HD\,152843~b}
\newcommand{\planetc}{HD\,152843~c}

\title[HD152843 b \& c]{HD152843 b \& c: the masses and orbital periods of a sub-Neptune and a super-puff Neptune}

\author[Nicholson et al.]{
B. A. Nicholson,$^{1,2}$\thanks{E-mail: belinda.nicholson@unisq.edu.au}
S. Aigrain,$^{1}$
N. L. Eisner,$^{3,4}$ 
M. Cretignier,$^{1}$ 
O. Barrag\'an,$^{1}$
L. Kaye,$^{1}$
\newauthor
J. Taylor,$^{1}$
J. Owen,$^{5,6}$ 
A. Mortier,$^{7}$ 
L. Affer,$^{8}$
W. Boschin,$^{9,10,11}$
A. Collier Cameron,$^{12}$
\newauthor
M. Damasso,$^{13}$
L. Di Fabrizio,$^{9}$
V. DiTomasso,$^{14}$
X. Dumusque,$^{15}$
A. Gehdina,$^{9}$
A. Harutyunyan,$^{9}$
\newauthor
D. W. Latham,$^{14}$
M. Lopez-Morales,$^{14}$
V. Lorenzi,$^{9}$
A. F. Martínez Fiorenzano,$^{9}$
E. Molinari,$^{16}$
\newauthor
M. Pedani,$^{9}$
M. Pinamonti,$^{13}$
A. Sozzetti,$^{13}$
K. Rice,$^{17,18}$
\\
$^{1}$Sub-department of Astrophysics, University of Oxford, Keble Rd, Oxford, United Kingdom, OX13RH\\
$^{2}$University of Southern Queensland, Centre for Astrophysics, Toowoomba, QLD, Australia, 4350\\
$^{3}$Center for Computational Astrophysics, Flatiron Institute, 162 Fifth Avenue, New York, NY 10010, USA \\
$^{4}$Department of Astrophysical Sciences, Princeton University, Princeton, NJ 08544, USA\\
$^{5}$Astrophysics Group, Department of Physics, Imperial College London, Prince Consort Rd, London SW7 2AZ, UK\\
$^{6}$Department of Earth, Planetary, and Space Sciences, University of California, Los Angeles, CA 90095, USA
$^{7}$School of Physics \& Astronomy, University of Birmingham, Edgbaston, Birmingham, B15 2TT, UK\\
$^{8}$INAF - Osservatorio Astronomico di Palermo, Piazza del Parlamento 1, 90134, Palermo, Italy\\
$^{9}$Fundacion Galileo Galilei INAF (Telescopio Nazionale Galileo), Rambla Jose Ana Fernandez Perez 7, 38712 Bren\~a Baja (La Palma), Canary Islands, Spain\\
$^{10}$ Instituto de Astrofisica de Canarias, C/Via Lactea s/n, 38205 La Laguna (Tenerife), Canary Islands, Spain\\
$^{11}$ Departamento de Astrofisica, Univ. de La Laguna, Av. del Astrofisico Francisco Sanchez s/n, 38205 La Laguna (Tenerife), Canary Islands, Spain\\
$^{12}$Centre for Exoplanet Science / SUPA, School of Physics \& Astronomy,University of St Andrews, North Haugh St Andrews, Fife, KY16 9SS, UK\\
$^{13}$ INAF - Osservatorio Astrofisico di Torino, Strada Osservatorio, 20, I-10025 Pino Torinese (TO), Italy\\
$^{14}$ Center for Astrophysics | Harvard \& Smithsonian, 60 Garden Street, Cambridge, MA 02138, USA\\
$^{15}$Observatoire de Geneve, 51 Chemin Pegasi, CH-1290, Versoix, Switzerland\\
$^{16}$ INAF - Osservatorio Astronomico di Brera \& REM via Bianchi 46, 23807 Merate (LC), Italy\\
$^{17}$Institute for Astronomy, University of Edinburgh, Royal Observatory, Blackford Hill, Edinburgh, EH9 3HJ, UK\\
$^{18}$Centre for Exoplanet Science, University of Edinburgh, Edinburgh, EH9 3HJ, UK
}

\date{Accepted XXX. Received YYY; in original form ZZZ}

\pubyear{2023}

\begin{document}
\label{firstpage}
\pagerange{\pageref{firstpage}--\pageref{lastpage}}
\maketitle

\begin{abstract}
We present the characterisation of the two transiting planets around \target\ (TOI\,2319, TIC\,349488688) using an intensive campaign of HARPS-N radial velocities, and two sectors of \tess\ data. These data reveal a unique and fascinating system: \target\ b and c have near equal masses of around 9 $M_{\oplus}$ but differing radii of \rpb and \rpc, respectively, and orbital periods of \Pb and \Pc days. This indicate that \planetc is in the lowest fifth-percentile in density of the known exoplanet population, and has the longest orbital period among these low density planets. Further, \planetc's radius places it in the `Saturn valley', the observed lack of planets larger than Neptune, but smaller than Saturn. The orbital periods of these planets indicate they are near a $5:3$ mean motion resonance, indicating the possibility of transit timing variations, and hints at the possibility of interaction with a third planet at some point in the evolution of this system. Further, the brightness of the host star and the low density of \planetc\ make it a key target for atmospheric characterisation. 
\end{abstract}




\section{Introduction}
Transiting exoplanets with well characterised orbits, masses and radii are highly valuable for detailed studies of planetary astrophysics. Even more valuable are such systems hosting multiple transiting planets - they provide robust constraints on planet formation and evolution models since these models must be able to explain the properties of all planets in the system. 

In the era of large-scale stellar photometric surveys searching for transiting planets, many multiple transiting planet systems have been found. However, many of these systems have been found around stars for which the follow-up radial velocity observations to measure their masses and full orbital characteristics are challenging: faint stars cannot provide the signal required for high-resolution spectroscopy, and the signal of a planet that is easy to detect in transit can be small in RV.    

\citet[][]{Eisner2021} (hereafter E21) presented the discovery of a system ideal for precise characterisation: two planets orbiting \target, a  bright (V = 8.85 mag), solar-type ($\mathrm{T_{eff}} = 6310 \pm 100$ K), main-sequence star. Transits of these planets, two of planet b and one of planet c, were seen in Sector 25 of NASA's Transiting Exoplanet Survey Satellite \citep[\tess,][]{Ricker15} primary mission and flagged by citizen scientists through the Planet Hunters \tess\ project \citep{eisner2020method}. E21 made robust radii measurements of both planets, measured the orbital period of planet b and placed constraints on the orbital period of planet c given its transit duration and the gaps and time span of the \tess\ data. Some radial velocity data were taken to validate the system, but these were not sufficient to provide robust mass measurements for either planet. 

Even with this limited information, \target\ stood out as an intriguing system: planet c lies in a sparsely populated region of the radius-period diagram, in the "gap" between Neptunes and Saturns / Jupiters ($5 < R_{\rm p} < 8$ \REarth, see Figure 9 of E21). This region contains both relatively high-density, warm sub-Saturns with a significant solid core, and extremely low-density `super-puffs'. The former case represents planets that have crossed the core mass threshold for runaway accretion towards a Jupiter-size, but have failed to do so due to the environment of the disc \citep[see e.g.:][]{Helled2023}. The nature and origins of the latter `super-puff' case are still uncertain. One possible scenario is stellar radiation creating high-altitude hazes that increase opacity out to larger planetary radii, making the planet appear larger than it would if its atmosphere was free of aerosols \citep{Lammer2016, GaoZhang2020}. Another possible scenario is that super-puff planets have additional sources of heating (i.e. tidal heating), that give rise to a thermally inflated radius \citep{Millholland2020}. 

Planet c is also intriguing as it resides in a system alongside a relatively `normal' sub-Neptune, planet b. Comparing and contrasting the properties of the two planets could yield useful insights into the processes that sculpt the "sub-Saturn valley" and/or the origin of super-puffs. However, precise constraints on the period and mass of planet c are needed in order to constrain theoretical models of how these planets are formed. 

This paper presents the precise masses and orbital parameters of \planetb\ and c, achieved through intensive RV monitoring with the HARPS-N instrument through the HARPS-N GTO program, and additional \tess\ data. In Section \ref{sec:obs}, we detail these observations and data reduction and processing steps. Section \ref{sec:host_star} describes the properties and parameters of the host star used to interpret our data, and explores any signs of stellar activity in the HARPS-N spectra and radial velocities. We show the results of the joint modelling of the transit and radial velocity data in Section \ref{sec:pyaneti} to obtain the orbital periods, masses and radii for both planets. With these parameters established, Section \ref{sec:discussion} discuss the implications of these on scenarios of formation and evolution of this system. We summarise and conclude our findings in Section \ref{sec:sum_concl}.

\section{Observations}
\label{sec:obs}

This work uses multi-year, high-precision, space-based photometry and ground-based radial velocity observation, spanning just over three years in both cases. While the observations from the first season were discussed in the discovery paper (E21), in this work, we re-analysed the full light curve and RV data sets and thus describe both in their entirety. 

\subsection{\tess\ photometry}

\tess\ first observed \target\ in Sector 25 of its primary mission, covering two transits of \planetb\ and one of c (E21). \target\ was re-observed by \tess\ in Sector 52 of its extended mission, during which two additional transits of planet b, and one of c were observed. 

\begin{figure*} 
    \centering
    \includegraphics[width=.9\linewidth]{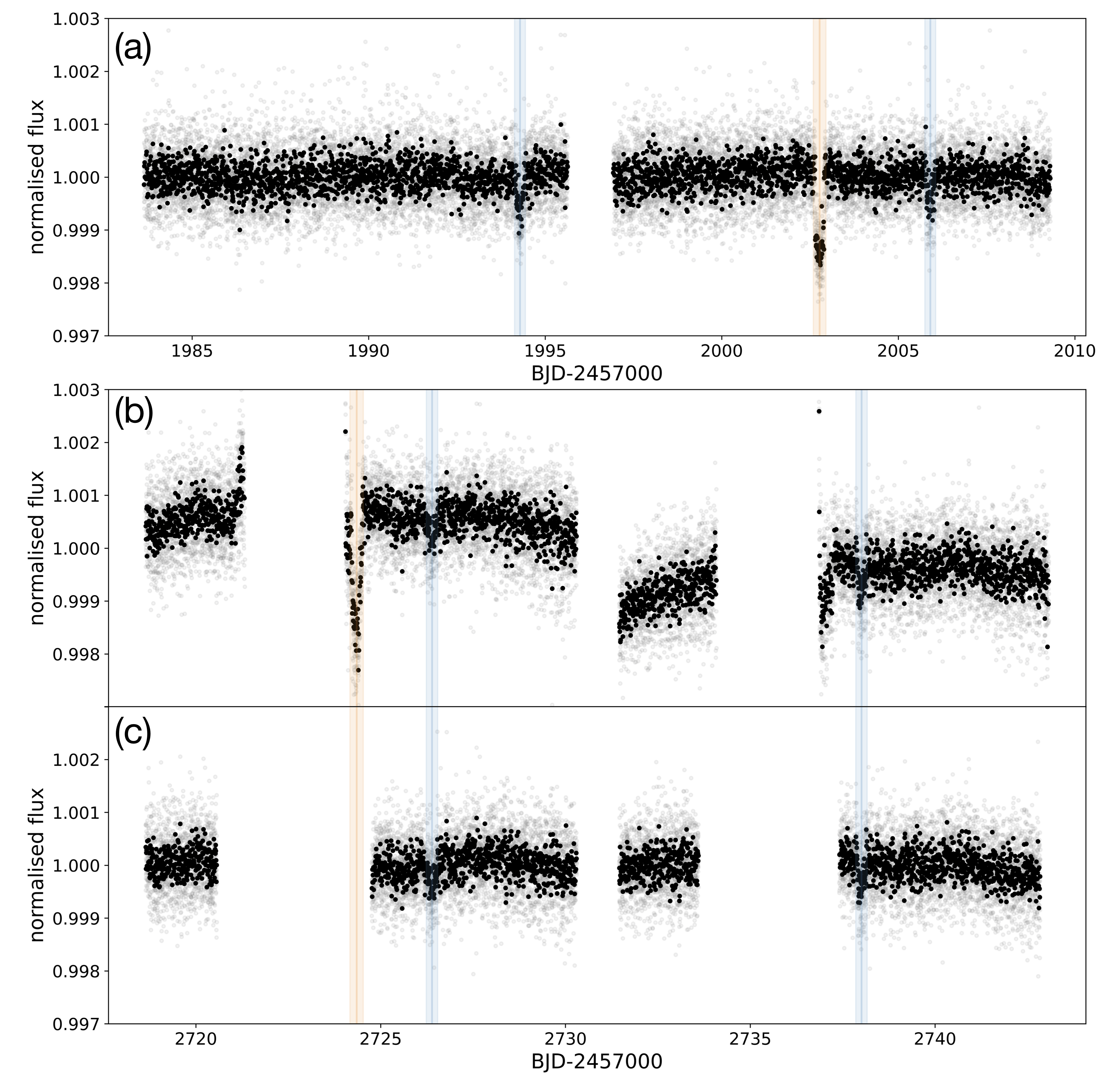}
    \caption{Normalised photometric \tess\ data for for \target. (a) Sector 25 PDC-SAP, (b) Sector 52 SAP, (c) Sector 52 PDC-SAP. In all panels, the grey points show the 2-min cadence data, the black points show the data binned to 10-min cadence, and the shaded blue and orange regions highlight transits of planets b and c, respectively.}
    \label{fig:LC_full}
\end{figure*}

In both sectors, the spacecraft obtained images at a cadence of 2\,seconds, which were combined onboard the satellite into 2\,minutes cadence data products. These were processed and reduced by the Science Processing Operations Center \citep[SPOC;][]{Jenkins2016}. Throughout this work, we use the pre-search data conditioning simple aperture photometry (PDC) light curve from the SPOC pipeline as our starting point. Sector 52 was affected by significant levels of Earth-shine \citep{Sector52Release}, which saturates portions of the detector and makes some sections of the light curve unusable, while introducing additional systematics in the rest. Parts of the light curve that were affected by intermediate levels of Earth-shine are included in the simple aperture photometry (SAP) light curve, but not in the PDC version. The second transit of \planetc\ falls within one of these segments. Since this transit is critical to pin down the period of \planetc, we processed the relevant section of the SAP light curve ourselves.


To process the Sector 52 SAP light curve for the transit of planet c, we used the {\sc Lightkurve} python package \citep{Lightkurve2018} to subtract the background flux, and correct for crowding and aperture losses (as is normally done as part of the PDC pipeline). To ascertain the potential impact of \tess\ systematics on the light curve around the transit of \planetc, we modelled the systematics using the Co-trending Basis Vectors (CBVs). These are computed by the PDC pipeline, and so were not available for the time-range of planet c's transit, but modelling the CBVs that were available allowed us to check that there were no large systematics around the transit, which would alter its shape or measured depth. 

The full \tess\ light curve is shown in Figure~\ref{fig:LC_full}, normalised to its median in each sector. The top panel shows Sector 25, including the two transits of \planetb\ (blue shading) and one transit of \planetc\ (orange shading), which were already analysed by E21. This light curve has a total of 17245 observations between 14th May and 8th June 2020, with a mean uncertainty of 5.0 ppt. The bottom panel shows the PDC light curve for sector 52, which has a total of 10843 observations between 19th May and 12th June 2022 with an average uncertainty of 4.5 ppt, and contains two additional transits of \planetb. The middle panel shows the SAP light curve for sector 52, including the section around the transit of \planetc, which was missing from the PDC light curve and has an additional 590 observations with an average uncertainty of 6.0 ppt. The $\sim 1$\,d data gap in the middle of each sector corresponds to the time taken for the spacecraft to point its antenna towards Earth (at the periapse of \tess's orbit), to send data to Earth and re-point. Additional data gaps in Sector 52 are caused by Earth-shine.



\begin{figure} 
    \centering
    \includegraphics[width=.9\linewidth]{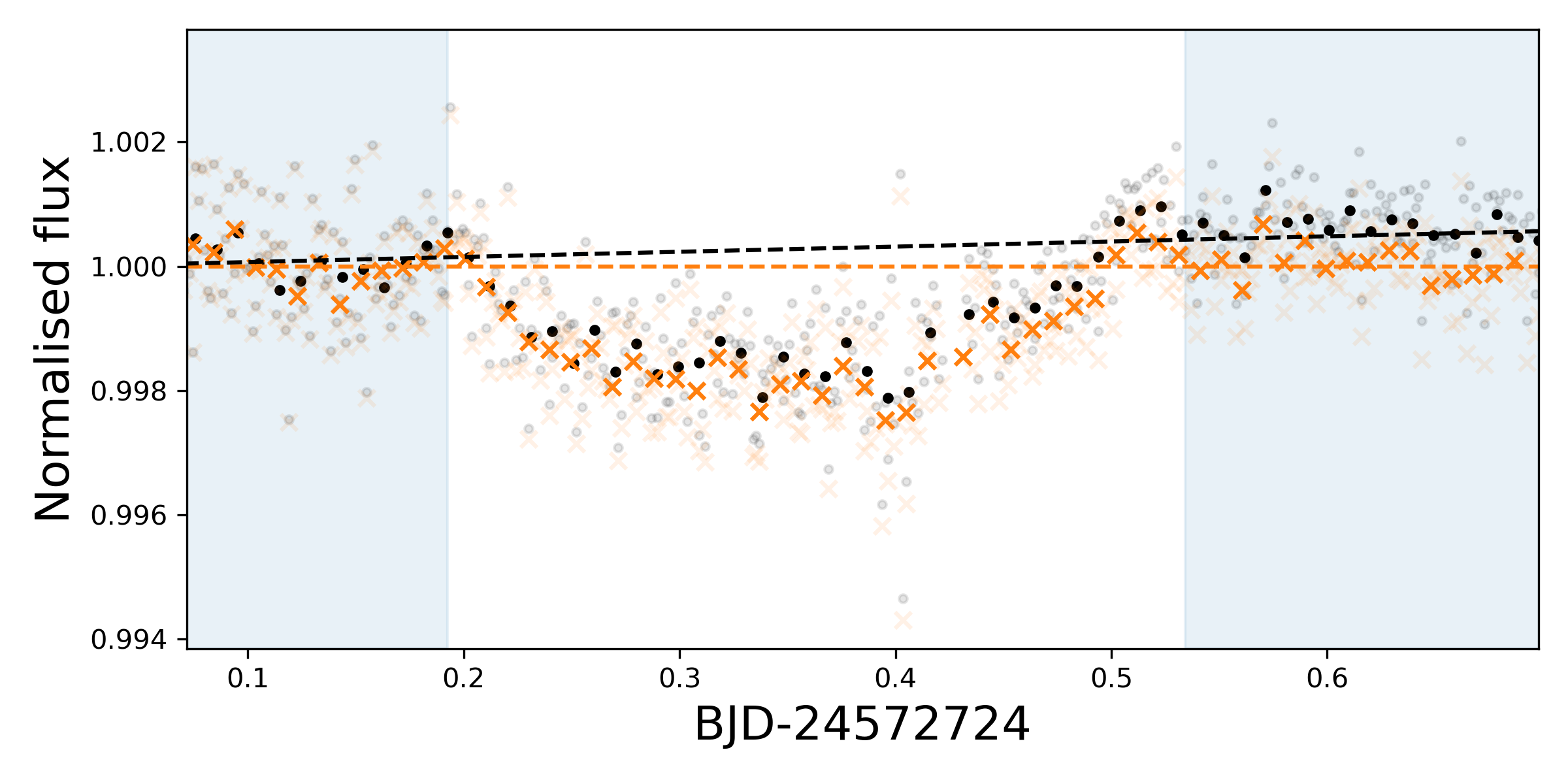}
    \caption{Fit to the out-of-transit baseline for the $2^{\text{nd}}$ transit of planet c (Sector 52, SAP data). The time-ranges used for the fit are indicated by the blue shaded areas. The input 2-min cadence data are shown by the grey dots and the resulting, normalised data by the faint orange crosses. The same data sets are also shown binned to 14-min cadence (black dots and orange crosses, respectively).}
    \label{fig:LC_c_52}
\end{figure}

To prepare the light curve for analysis, we select segments of the light curve lasting 0.5\,d around each transit. We used the PDC light curve wherever available, since that is expected to be the most reliable, and used our processing of the SAP light curve only for the second transit of planet c. Due to the gap in the \tess\ data and significant Earth-shine prior to the transit of \planetc, the out-of-transit baseline prior to this transit is shorter than the 0.5\,d cutoff. We normalise the baseline flux around each transit by fitting a first order polynomial to the out of transit flux and dividing by the result. This process is illustrated in Figure~\ref{fig:LC_c_52} for the second transit of planet c (Sector 52, SAP data). This is the transit with the steepest baseline trend: the gradient of the light curve around the transit is significantly smaller for all the other transits, where we used the PDC data. 

\subsection{HARPS-N spectroscopy}

We acquired high-resolution ($R\approx 115\,000$) spectra with the High Accuracy Radial velocity Planet Searcher in the Northern hemisphere \citep[HARPS-N;][]{Cosentino2012, Cosentino2014} spectrograph mounted at the 3.6-m Telescopio Nazionale Galileo in La Palma, Spain, via a Director’s Discretionary Time program (ID: A41DDT4) and subsequently through the HARPSN-GTO program. We obtained 18 spectra between 5 September and 11 November 2020, 65 spectra between 3 March and 23 September 2021, and 87 spectra between 4 February and 5 October 2022. Most of exposures were at 900\,s, with a handful of longer exposure (1x1800\,s, 1x1500\,s and 4x1200\,s) to compensate for poor conditions, and two aborted observations that have exposure times too short to be useful. The resulting mean Signal-to-Noise Ratio (SNR) at 550\,nm of the completed observations is 101. The wavelength solution was determined by simultaneous wavelength calibration with a Fabry–Perot etalon and nightly calibration frames with a Thorium-Argon lamp. The first season of observations, analysed by E21, was originally reduced using the standard HARPS Data Reduction Software \citep[DRS version 3.7;][]{Baranne1996}, including RV extraction via cross-correlation with a G2 spectral template, yielding a mean RV uncertainty of $\sim 4.2\,\ms$. In this work, all the observations were re-reduced using the more recent version 2.3.5 of the DRS \citep{Dumusque2021}, which has been shown to reduce systematics. In addtition to calculating the radial velocities, the DRS also computes the moments and bisector span of the cross-correlation function (CCF), and an S-index and $\log R'_{\rm HK}$ from the Ca II H\&K lines as stellar activity indicators.

\subsubsection{Post-processing and radial velocity determination with {\sc YARARA}}

For moderate rotators such as \target\ ($\vsini=8.38\pm0.5\kms$), the cross-correlation RV extraction process yields larger RV uncertainties than that for slower rotators, as the broadening of the spectral lines reduces the accuracy with with the position of the line centre can be measured, and may make the spectral features more sensitive to instrumental systematics and telluric lines. Since the DRS spectra used here contain residual instrumental and telluric signatures, we attempt to improve the accuracy of our RV estimates by post-processing the time-series of DRS-reduced 1-D spectra using {\sc YARARA} \citep{Cretignier2021}.

The post-processing begins from the 1-D spectra produced by the official DRS \citep{Dumusque2021}. A subset of the observations (totalling 136) were selected, as {\sc YARARA} requires a homogeneous data set of high SNR observations. Specifically,  spectra with SNR$<75$ were excluded, as were those corresponding to $\ge 3 \sigma$ outliers in one of the CCF moments or the $S$-index produced by the DRS. These outliers typically indicate exposures that have been taken under sub-optimal observing conditions (e.g.\ clouds or moonlight) or that have been reduced using non-standard calibration frames. 

\begin{figure*} 
    \centering
    \includegraphics[width=\linewidth]{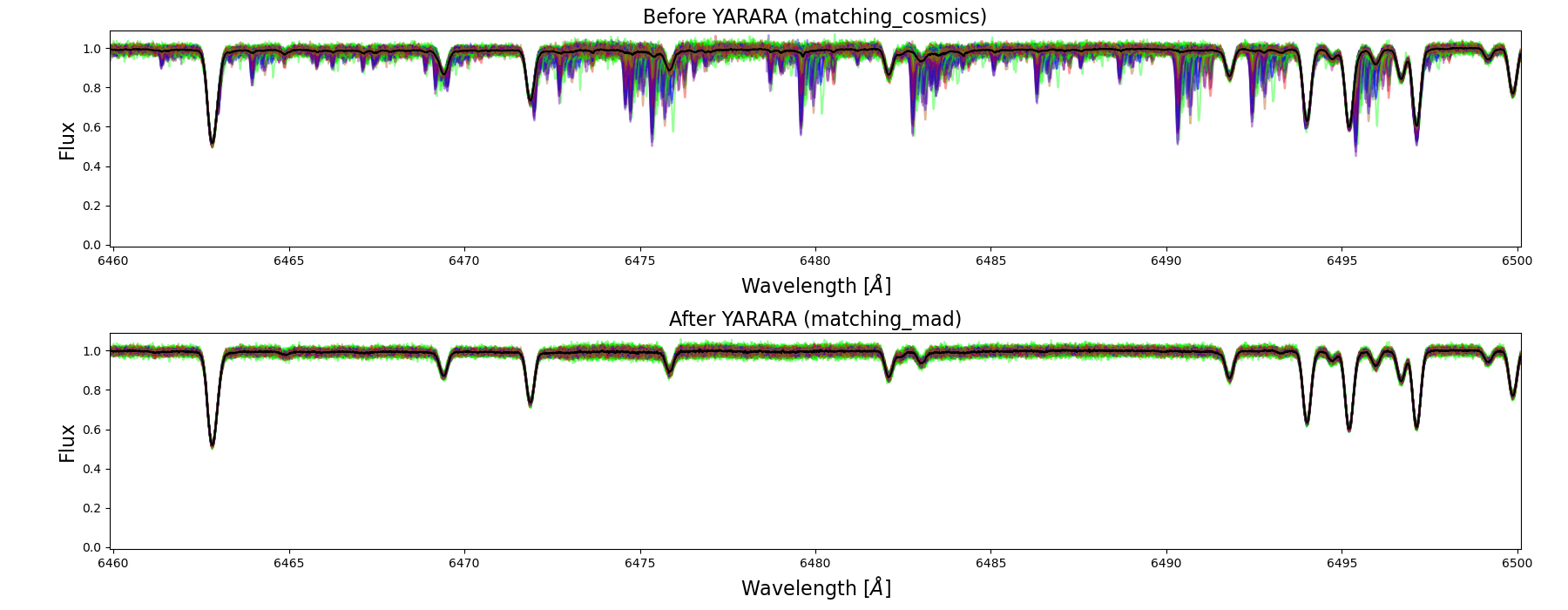}
    \caption{Correction of telluric lines by {\sc YARARA}. The individual spectra are each plotted in a different colour before (top) and after (bottom) {\sc YARARA} post-processing.}
    \label{fig:YARARA}
\end{figure*}

The remaining spectra were first continuum-normalised using a convex hull algorithm implemented in the {\sc Python} package {\sc RASSINE} \citep{Cretignier2020b}, which has been shown to approximate the continuum at a precision compatible with the photon-noise of the observations. The continuum-normalised spectra were then median-combined to construct a master stellar spectrum, which was subtracted from the individual spectra. The matrix of residual spectra was then analysed to reveal and correct for flux systematics such as: cosmic rays, tellurics, ghosts, interference patterns and point spread function (PSF) variations \citep{Cretignier2021}. {\sc YARARA} can also correct stellar activity effects in the spectra, but this step was not performed here, since the star is not active (see Section~\ref{sect.activity}). Even though the median SNR of the observations is only $\sim100$, whereas the performance of {\sc YARARA} is at its full potential when applied for bright targets (SNR>250), we still see a clear improvement in the spectra, as illustrated by Figure~\ref{fig:YARARA} for a region contaminated by telluric lines. 

The DRS usually computes a CCF using a generic line list, optimised for a broad spectral type class. As shown in \citet{Bourrier2021}, this results in sub-optimal RV estimates, and improved RVs can be obtained by deriving a bespoke line list, tailored for the star's specific temperature and metallicity, from the spectra themselves following the method of \citet{Cretignier2020a}. Here, we followed the same procedure except that we did not exclude lines contaminated by telluric lines, since those are now corrected by {\sc YARARA}. Once the CCF is computed, the RV extraction proceeds by fitting a Gaussian to the CCF as in the standard DRS.

\subsection{Final RVs and activity indicators}
Table~\ref{tab:RVs} summarises the full set of observations of \target\ from HARPS-N. The full list is included as supplementary material to this paper. For the analysis that follows we use the YARARA radial velocities, and activity indicators. The median RV uncertainty and SNR of this final data set are 1.71\,\ms and 108, respectively. For reference, the exposure times for each observation, and the main pipeline radial velocities are included, and a flag column indicating if an observation is included in the final analysis. In addition to excluding poor SNR or incomplete observations, we also exclude observations taken between the $8^{\text{th}}$ and the $12^{\text{th}}$ of May 2021 from the analysis, due to a known problem with the guiding camera on those nights (private communication, TNG). While the observations of \target\ are not visibly affected, standard stars observed on those nights display an offset in their RV measurements, which undermined our confidence in the RVs for our target on those dates.

\begin{table*}
\centering
  \caption{Journal of observations of HD152843 taken with the HARPS-N instrument, including signal to noise ratio (SNR) calculated at 550\,nm, standard pipeline (DRS) radial velocity measurements with uncertainties, as well as radial velocities and activity indicators computed by the {\sc Yarara} pipeline for observations included in the analysis, indicated with `Use Flag'=1. Some observations were excluded from the analysis (`Use Flag'=0) due to low signal to noise (SNR < 75) or outliers rejection criterion (see main text). The full table is provided in machine-readable format as supplementary material. \label{tab:RVs}}  
  \begin{tabular}{ccccccccccc}
  \hline
	Time		 & SNR & \multicolumn{2}{c}{DRS} & \multicolumn{6}{c}{{\sc Yarara}}                                                                                           & Use Flag \\
 \cmidrule(rl){3-4} \cmidrule(rl){5-10}
	(BJD-2457000)&             & RV			&	$\sigma_{RV}$	& RV			&	$\sigma_{RV}$	& $\log R'_{HK}$ & $\sigma_{\log R'_{HK}}$ & Bis. Span & $\sigma_{\rm Bis. Span}$ & 1 or 0   \\
	             &             & (\ms) 		&	(\ms)		    & (\ms) 		&	(\ms)		    &                &                         &               &                              &  \\
   \hline
2098.3530 &       157 &  9872.394874 &   1.691244 &  2.575514 &   1.246248 & -5.308577 &  0.452185 &      9.453929 &          2.245117 &      1 \\
2102.3420 &       120 &  9863.510576 &   2.268848 & -4.495355 &   1.467149 & -5.329140 &  0.456954 &    -52.470551 &          3.909014 &      1 \\
2104.3659 &        20 &  9869.671016 &  16.359800 &       NaN &        NaN &       NaN &       NaN &           NaN &               NaN &      0 \\
2110.3261 &        87 &  9864.607312 &   3.144023 & -4.607612 &   3.189392 & -5.149874 &  0.463961 &     -2.009999 &          2.704267 &      1 \\
2111.3796 &       100 &  9868.244643 &   2.712388 &  2.353805 &   1.714082 & -5.748914 &  0.458733 &    129.392928 &          3.923864 &      1 \\
2114.3398 &        67 &  9876.167603 &   4.003329 &       NaN &        NaN &       NaN &       NaN &           NaN &               NaN &      0 \\
    \vdots    & \vdots & \vdots  & \vdots  & \vdots  & \vdots  & \vdots  & \vdots  & \vdots  & \vdots  & \vdots \\
    \hline
   \noalign{\smallskip}
  \end{tabular}
\end{table*}

\section{Parameters and activity of the host star}
\label{sec:host_star}

We use the fundamental stellar parameters determined by E21 using the first 18 HARPS-N spectra, which are summarised in Table \ref{tab:stellar_pars}. The uncertainties on those parameters are dominated by systematic rather than random errors, and therefore re-estimating the parameters using the more recent HARPS-N observations would not yield significantly different values. 

Given the significantly larger number of spectra, we can however investigate the activity of the host star in more depth than was done by E21.

\begin{table}
    \centering
    \begin{tabular}{lc}
    \hline
    Effective Temperature $\mathrm{T_{eff}}$ (K)  & {$6310 \pm 100 $ }              \\
    Surface gravity $\log g_\star$ (cgs)          & {$4.19 \pm 0.03 $ }             \\ 
    \vsini (\kms)                                 & {$8.38 \pm 0.50$  }  \\
    $\mathrm{[M/H]}$ (dex)                                 & {$-0.22 \pm 0.08$ }  \\
    $\mathrm{[Fe/H]}$ (dex)                                & {$-0.16 \pm 0.05$ }  \\
    $v_{\rm mic}$ (\kms)                          & {$1.66 \pm 0.13$  }  \\
    $v_{\rm mac}$ (\kms)                          & {$\sim2$} \\
    Stellar mass $M_{\star}$ ($M_\odot$)          & {$1.15 \pm 0.04 $ }  \\
    Stellar radius $R_{\star}$ ($R_\odot$)        & {$1.43 \pm 0.02 $ }  \\
    Stellar density $\rho_\star$ ($\rho_\odot$)   & {$0.40 \pm 0.03 $ }  \\
    Star age (Gyr)                                & {$3.97 \pm 0.75$  } \\
    \hline
    \end{tabular}
    \caption{Stellar parameters of HD152843 used for the analysis in this paper. Values are from E21, except where noted: [add note] }
    \label{tab:stellar_pars}
\end{table}

\subsection{Activity analysis}
\label{sect.activity}

Our $\log R'_{\rm HK}$ values for \target\ have a roughly Gaussian distribution around a value of $-5.25$ with a standard deviation of 0.06, except for a single outlier at -5.70. These low values suggests that \target\ is a quiet star, and that activity is not expected to affect the RVs significantly. 

\begin{figure} 
    \centering
    \includegraphics[width=\linewidth]{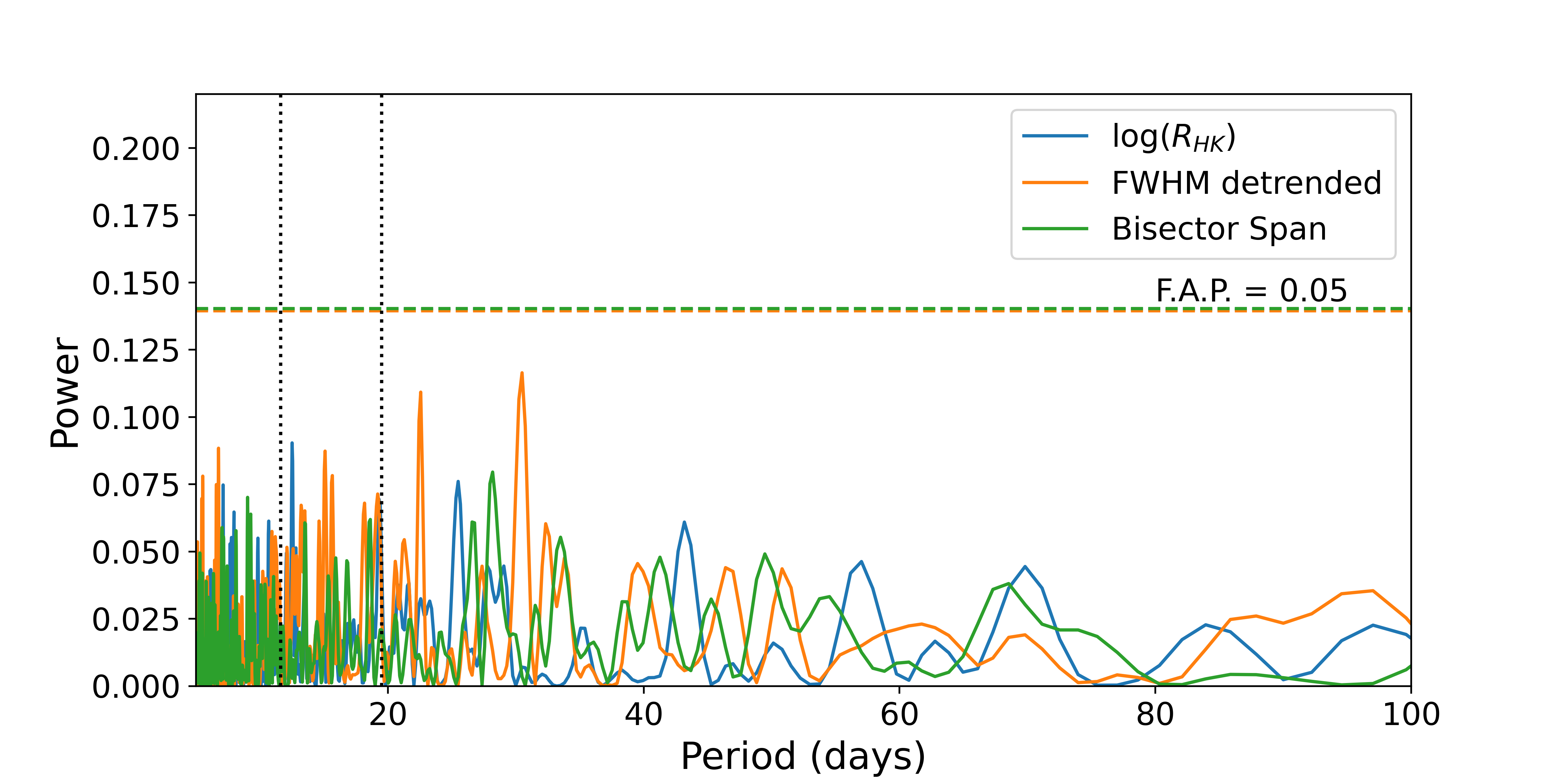}
    \caption{Lomb-Scargle periodograms of the $\log R'_{\rm HK}$ (blue), detrended FWHM (orange) and bisector span (${\rm BIS}$, green) stellar activity indicators. The 5\% false alarm probability level for each periodogram is indicated by the horizontal dashed line of the same colour. }
    \label{fig:act_PGram}
\end{figure}

To double-check this, we computed the Pearson correlation coefficient between the $\log R'_{\rm HK}$ values and the RVs, which is $0.15$, indicating very little correlation. We also computed Lomb-Scargle periodograms\footnote{All Lomb-Scargle periodograms were computed in {\sc Python} using the {\sc time\_series.LombScargle} module of the {\sc astropy} package.} of the chromospheric and CCF-based activity indicators, which are shown in Figure~\ref{fig:act_PGram}. In the case of the CCF Full Width at Half Maximum (FWHM), a small linear trend, which may be due to instrumental effects (focus drift) or secular changes in activity level (activity cycle), was subtracted before computing the periodogram. None of the indicators display  peaks in their periodogram above the false alarm probability (FAP) threshold for significance of 5\%, and so show no significant periodicity. We place an upper limit on the stellar rotation period of $\approx 8.6$\,d based on the measured and estimated values of \vsini\ and $R_\star$. 

Finally, the light curve shown in \ref{fig:LC_full} shows no signs of spot-induced variability. We therefore proceed to model the light curve and RVs without accounting for activity effects, which should be negligible compared to other sources of uncertainty.

\section{Joint modelling of the transits and RVs}
\label{sec:pyaneti}

In this section, we describe the joint analysis of the \tess\ light curve and HARPS-N RVs to derive the parameters of the two planets.

\subsection{Periodogram analysis of the RVs}

\begin{figure} 
    \centering
    \includegraphics[width=\linewidth]{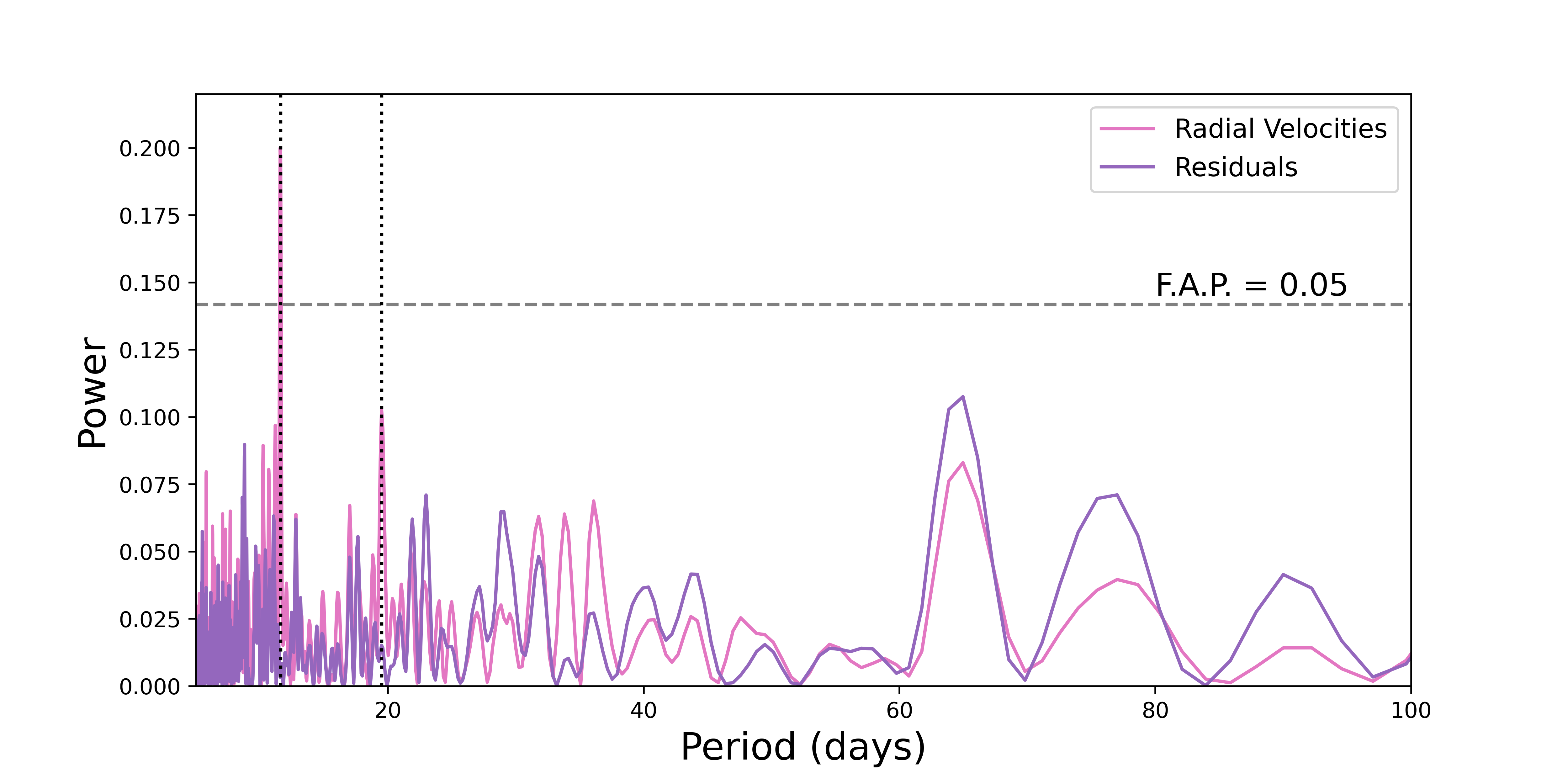}
    \caption{Lomb-Scargle periodograms of the raw RVs (pink) and the residuals of the two-planet fit (purple).}
    \label{fig:RV_pgram}
\end{figure}

\begin{figure} 
    \center
    \includegraphics[width=\linewidth]{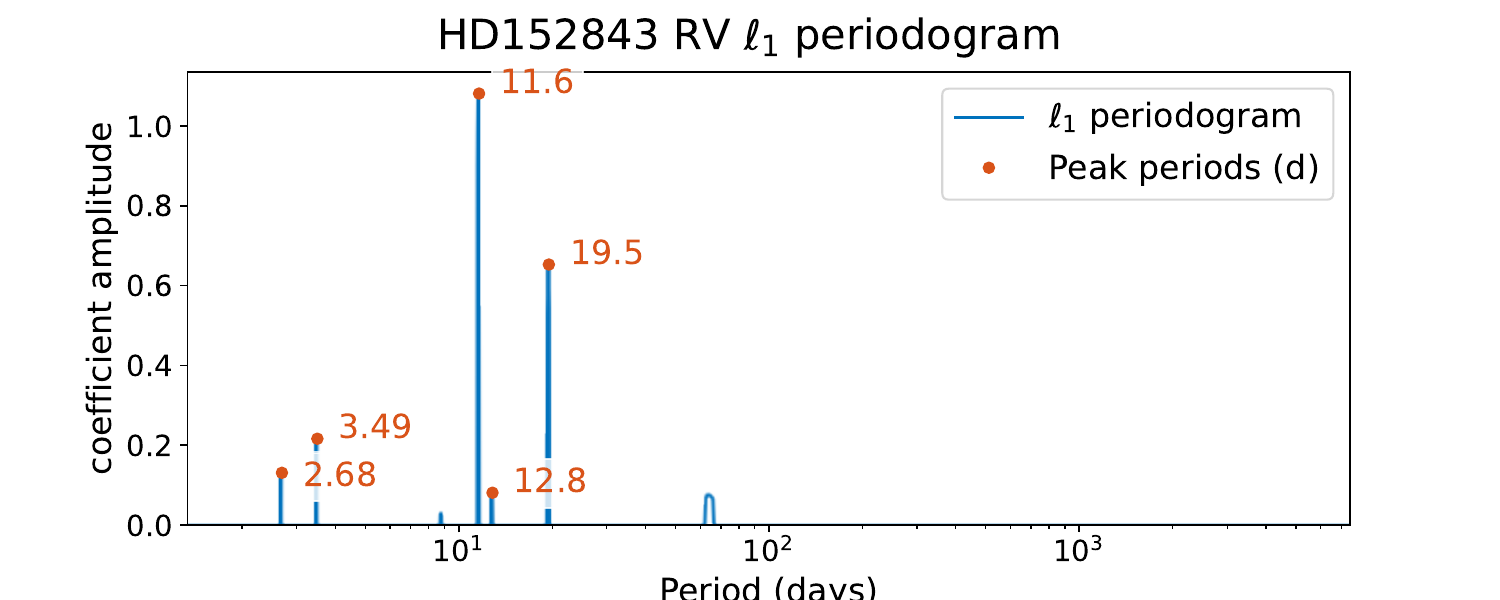}
    \caption{$l_1$ periodogram of the radial velocity time series of \target. The two highest peaks have FAP values of 0.4$\%$ and 2.1$\%$, and correspond to the orbital periods of planet b and c.}
    \label{fig:L1_pgram}
\end{figure}

As a starting point of this analysis, we computed the Lomb-Scargle periodogram of the raw RVs check if the RV signals of the two transiting planets are present in the RVs. This is shown as the pink line in Figure~\ref{fig:RV_pgram}. Peaks at the periods of both planets are clearly visible, though only the peak corresponding to planet b is statistically significant in this periodogram, which assumes the data consists of a single sinusoidal signal plus white noise. We therefore also computed the L1 periodogram \citep{Hara(2017)}, which is designed to search for signals from multiple planets at the same time. The periodogram is shown in Figure~\ref{fig:L1_pgram}. There are only two significant peaks, with (${\rm FAP} < 5\%$), corresponding to the periods of 11.62 days 19.50 days, which correspond to the measured and minimum possible orbital periods of planets b and c from \tess\ transit photometry. 

Having established that the signals of both planets are present in the RVs, we now proceed to model the photometry and the RVs jointly.

\subsection{Method and priors}
\label{sec:pyaneti_setup}

The transit and RV data were jointly analysed using the open access {\sc Python} package {\sc pyaneti} \citep[][]{pyaneti1, pyaneti2}.
In brief, {\sc pyaneti} fits the normalised transits using the model of \citet{Mandel2002} with quadratic limb-darkening, and the RVs using Keplerian orbits, and samples the joint posterior over the parameters of the transit and RV fit using a  Markov Chain Monte Carlo (MCMC) approach. All fitted parameters and priors used for the joint modelling are presented in Table~\ref{tab:parstarget}. We also fit for the stellar density $\rho_\star$, and recover the scaled semi-major axis for each planet in the system using Kepler's third law. We use a Gaussian prior on $\rho_\star$ based on the stellar mass and radius derived by E21 and reported in Table~\ref{tab:parstarget}. We allowed for eccentric orbits for both planets, since their periods are too long to assume that the orbits have been tidally circularised.

For all our {\sc pyaneti} runs, we sampled the parameter space using MCMC with 500 independent chains and created posterior distributions using 5000 iterations of converged chains with a thin factor of 10. Convergence was measured by the convergence test method of \cite{Gelman1992}, where chains are considered converged when the scaled potential factor, which is function of the variance within and between chains, is less than 1.02 \citep{Gelman2003}.  Overall, this resulted in $250\,000$ independent samples for each parameter.

\begin{table*}
\centering
  \caption{Parameters of the joint transit and RV fit. \label{tab:parstarget}}  
  \begin{tabular}{lcc}
  \hline
  Parameter & Prior$^{(\mathrm{a})}$ & Value$^{(\mathrm{b})}$\\
  \hline
  \multicolumn{3}{l}{\textit{\textbf{Model Parameters for \target b}}} \\
  \noalign{\smallskip}
    Orbital period $P_{\mathrm{orb}}$ (days)  &  $\mathcal{U}[11.620,11.622]$ & \Pb[] \\
    & & \\
    Transit epoch $T_0$ (BJD - 2457000)  & $\mathcal{U}[1994.25 , 1994.30]$ & \Tzerob[]  \\
    & & \\
    Parametrization $e \sin \omega$  &  $\mathcal{U}[-1,1]$ & \esinb[]  \\
    & & \\
    Parametrization $e \cos \omega$   &  $\mathcal{U}[-1,1]$ &\ecosb[] \\
    & & \\
    Scaled planet radius  $R_\mathrm{p}/R_{\star}$ &  $\mathcal{U}[0,0.1]$ & \rrb[]  \\
    & & \\
    Impact parameter, $b$ &  $\mathcal{U}[0,1.1]$  & \bb[] \\
    & & \\
    Doppler semi-amplitude, $K$ (\ms) & $\mathcal{U}[0,50]$ & \kb[]\\
    \noalign{\smallskip}
    \multicolumn{3}{l}{\textit{\textbf{Model Parameters for \target c}}} \\
    \noalign{\smallskip}
    Orbital period $P_{\mathrm{orb}}$ (days)  &  $\mathcal{U}[19.499,19.505]$ & \Pc[] \\
    & & \\
    Transit epoch $T_0$ (BJD - 2457000)  & $\mathcal{U}[ 2002.76 , 2002.78]$ & \Tzeroc[]  \\
    & & \\
    Parametrization $e \sin \omega$  &  $\mathcal{U}[-1,1]$ &  \esinc\\
    & & \\
    Parametrization $e \cos \omega$   &  $\mathcal{U}[-1,1]$ & \ecosc \\
    & & \\
    Scaled planet radius  $R_\mathrm{p}/R_{\star}$ &  $\mathcal{U}[0,0.1]$ & \rrc[]  \\
    & & \\
    Impact parameter, $b$ &  $\mathcal{U}[0,1.1]$  & \bc[] \\
    & & \\
    Doppler semi-amplitude, $K$ (\ms) & $\mathcal{U}[0,50]$ & \kc[] \\
    \multicolumn{3}{l}{\textit{\textbf{Other Parameters}}} \\
    \noalign{\smallskip}
    Stellar density $\rho_\star$ (${\rm g\,cm^{-3}}$) &  $\mathcal{N}[0.56,0.04]$ & \denstrb[] \\
    & & \\
    Parameterized limb-darkening coefficient $q_1$ $^{(\mathrm{c})}$ & $\mathcal{U}[0,1]$ & \qone \\
    & & \\
    Parameterized limb-darkening coefficient $q_2$ $^{(\mathrm{c})}$ & $\mathcal{U}[0,1]$ & \qtwo \\
    & & \\
    Offset velocity HARPS-N (\kms) & $\mathcal{U}[-0.50 , 0.50]$ & \HARPSN[] \\
    & & \\
    Jitter HARPS-N (\ms) & $\mathcal{U}[0,100]$ & \jHARPSN[] \\
    & & \\
    Jitter \tess\ (ppm) & $\mathcal{U}[0,500]$ & \jtr[] \\
    \hline
   \noalign{\smallskip}
  \end{tabular}
  \begin{tablenotes}\footnotesize
  \item \textit{Note} -- $^{(\mathrm{a})}$ $\mathcal{U}[a,b]$ refers to uniform priors between $a$ and $b$, $\mathcal{N}[a,b]$ to Gaussian priors with mean $a$ and standard deviation $b$.
  $^{(\mathrm{b})}$  The reported parameter values and errors are defined as the median and 68.3\% credible interval of the posterior distribution. 
  $^{(\mathrm{c})}$ $q_1$ and $q_2$ parameters as in \citet{Kipping2013} 
\end{tablenotes}
\end{table*}

\subsection{Determining the period of planet c}
\label{sec:p_c}

\begin{figure}
    \centering
    \includegraphics[width=\linewidth]{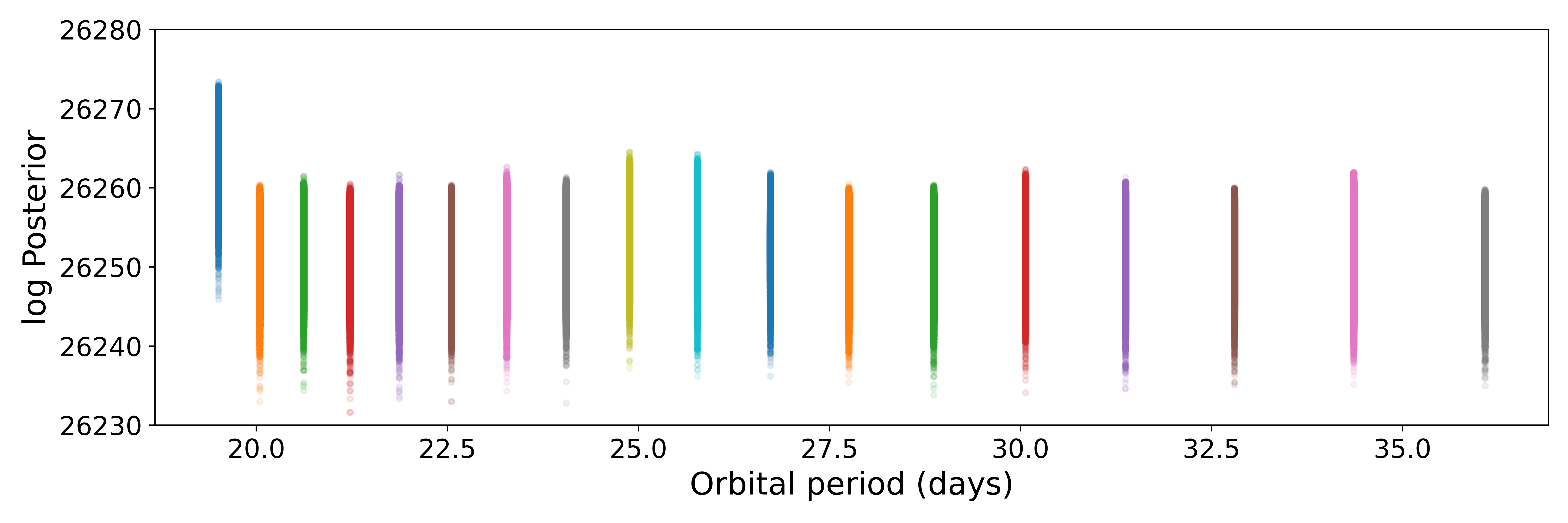}
    \caption{Log posterior for the period of \planetc\ over the 19.26--35\,d range allowed by the \tess\ light curve (see text for details). The shortest period is clearly favoured.}
    \label{fig:Pc_allperiods}
\end{figure}

The two transits we have observed of planet c are not sufficient to determine its orbital period, due to the large gap ($\sim 2$ years) between the two \tess\ sectors containing the transit events. We can, however, place constraints on what the possible periods might be from the \tess\ data, before refining this estimate using the RV time series. E21 constrained the minimum possible period to be 19.26 days from the data gaps and timing of the transit in the Sector 25 light curve, and this does not change with the additional \tess\ sector. They also derived an approximate upper limit on the period of $\sim 35$ days based on the transit duration and stellar parameters. Within this broad range, $P_c$ is constrained to lie within a number of narrow windows depending on the (integer) number of orbits separating the two observed transits. 

The RVs can be used to discriminate between the different possible periods for planet c. Initially, we attempted a joint fit to the transits and RV data using a broad uniform prior on $P_c$, spanning 19.26--35 days. However, the posterior for $P_c$ is highly multi-modal within that range, with over a dozen sharp, narrow peaks. The MCMC algorithm used by {\sc pyaneti} is designed to `hone in' on the strongest mode(s) of the joint posterior distribution, and consequently fails to explore all of these peaks properly when given a broad uniform prior. Since we know the approximate location of the individual peaks, however, we can work around this problem by sampling them individually. 

We therefore carried out separate {\sc pyaneti} runs with a narrow, uniform prior on $P_c$ around each possible period between 19.2 and 35 days given an integer number of orbits, with a width of 0.006\,days. All the other parameter priors were as described in Table~\ref{tab:parstarget}. The resulting log posterior distributions are plotted in Figure~\ref{fig:Pc_allperiods}. Because we used the same priors for each of these runs, changing only the location (but not the width) of the prior over $P_c$, the log posterior values for the samples from each run can be converted to log likelihoods by subtracting the same constant, and can be compared with each other directly. The marginal posterior distribution over $P_c$ (and hence the marginal likelihood function) for each peak is well approximated by a Gaussian, and the log likelihood for the shortest-period peak, at $P_c \sim 19.5$ days (corresponding to 37 orbits between the two observed transits) exceeds the peak log likelihood for all the other peaks by $>10$ (see Figure~\ref{fig:Pc_allperiods}). This indicates strong support for the shortest allowed period. Furthermore, this run is the one that yields the clearest detection of the RV signal of planet c (in terms of semi-amplitude over corresponding uncertainty), and is in agreement with the second most significant peak found in the $l1$ periodogram analysis of the RV data. We therefore use this value for $P_c$ in our final analysis, described in the next section.

\subsection{Final fit}
\label{sec:pyaneti_final}

\begin{figure} 
    \centering
    \includegraphics[width=\linewidth]{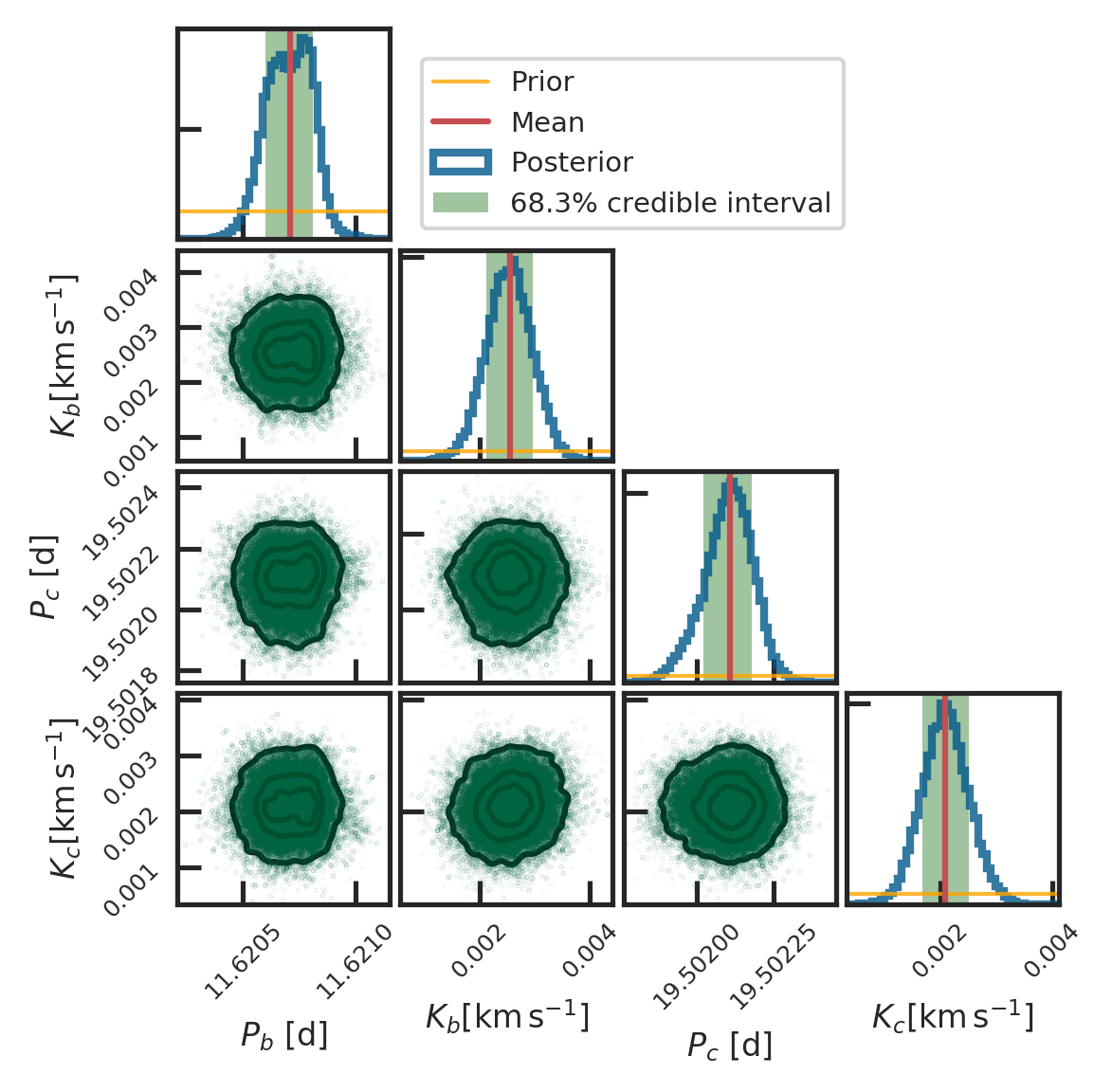}
    \caption{1- and 2-D posterior distributions for the orbital periods and semi-amplitudes of planets b and c from the final MCMC analysis. In the 1-D plots on the diagonal, the blue histograms show the distribution of the samples, the red vertical line shows the median, and the green shaded region shows the $\pm1\sigma$ interval. The horizontal orange line denotes the prior, which for all the parameters shown here was uniform and extends beyond the range of the plots. In the off-diagonal, 2-D parameter correlation plots, the green points show the individual samples and the contours show the 68\%, 95\% and 99\% confidence intervals.}
    \label{fig:idefix_correlations}
\end{figure}

\begin{figure*} 
    \centering
    \includegraphics[width=\linewidth]{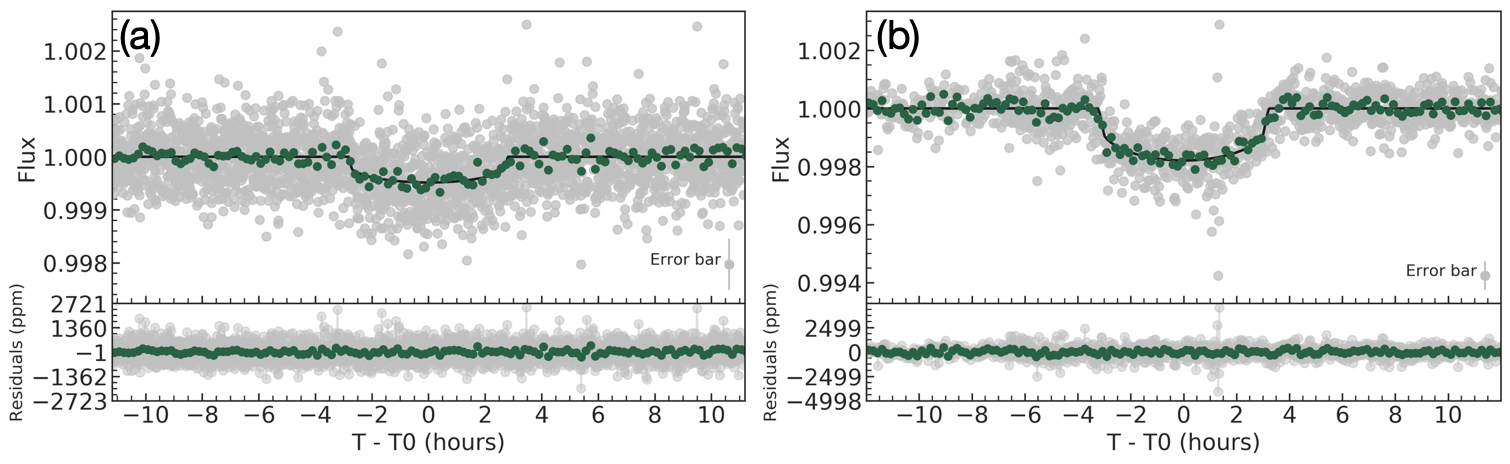}
    \caption{(a) \tess\ light curve phase-folded at the period of planet b, with the best-fit transit model shown by the black line, and the residuals of this model shown in the bottom panel. The 2-min cadence data used in the fit are shown as grey points, while the green points show the same data binned by a factor of 5 in phase. (b): same for planet c.}
    \label{fig:idefix_transits}
\end{figure*}

\begin{figure*} 
    \centering
    \includegraphics[width=\linewidth]{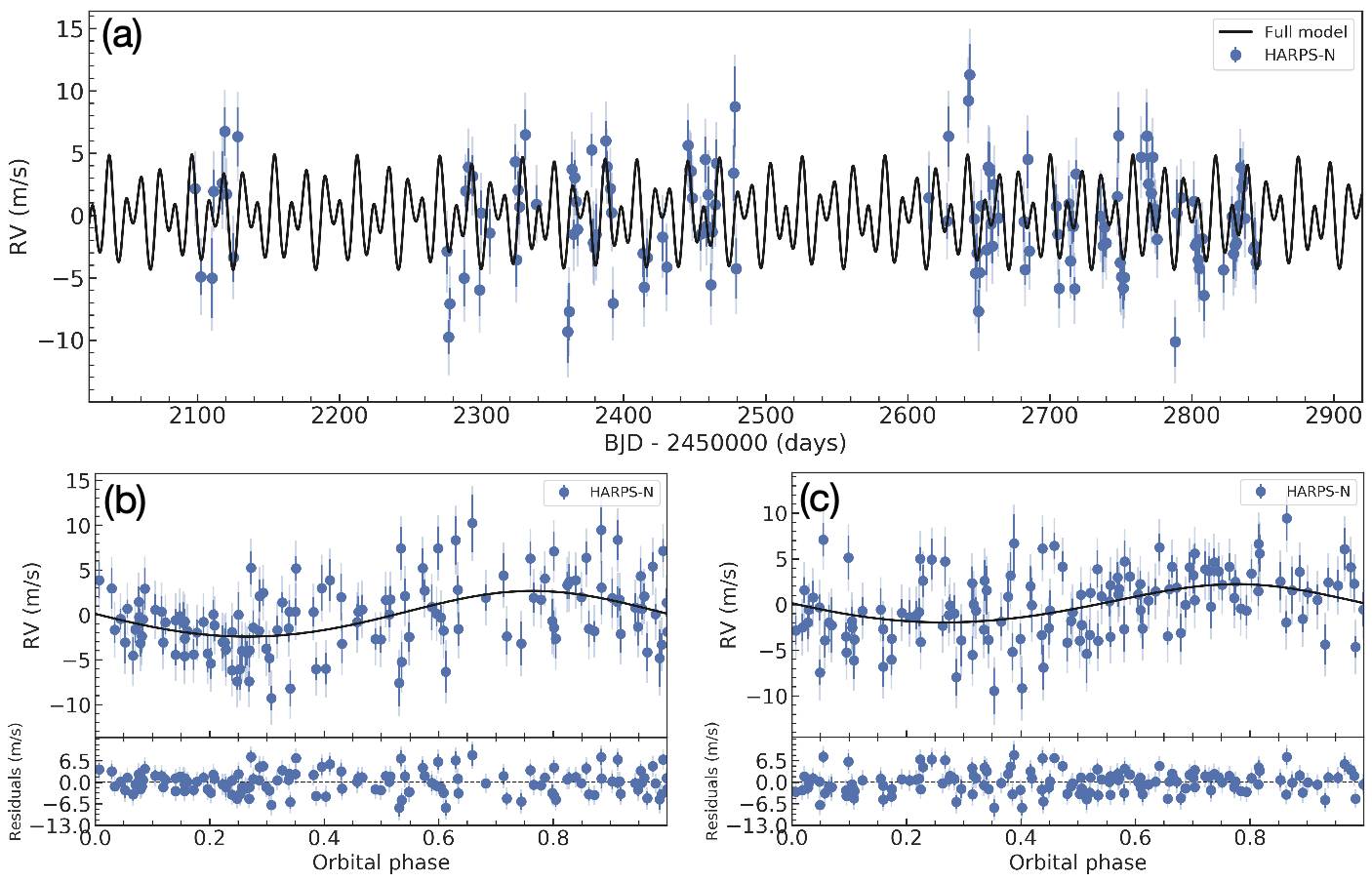}
    \caption{(a) Final fit (black line) to the HARPS-N RVs (blue points). (b) and (c) phase-folded RV curves for planets b and c, respectively, after subtracting the signal from the other planet, with the residuals shown in the lower panel. In all plots, the semi-transparent error bars indicate the jitter level added in quadrature to the formal uncertainties.}
    \label{fig:idefix_rvs}
\end{figure*}

Having ascertained that the shortest allowed period is the preferred one for planet c, we adopted the corresponding {\sc pyaneti} run, which set a uniform prior over $P_c$ in the range 19.499 days to 19.505 days, as the final one. The fitted parameters, extracted from the resulting posteriors, can be found in Table~\ref{tab:parstarget}. The 1- and 2-D posterior distributions for the orbital periods and semi-amplitudes of the two planets are shown in Figure~\ref{fig:idefix_correlations}, showing that the posteriors for these parameters are approximately Gaussian and that the RV signals of both planets are detected at the $\sim 5\sigma$ level. The fits to the transits and RV data are shown in Figures~\ref{fig:idefix_transits} and \ref{fig:idefix_rvs}, respectively. A figure showing the full set of 1- and 2-D posterior distributions is provided in supplementary material. 

The RV semi-amplitudes of both planets are non-zero at the $>5\sigma$ level (see Figure~\ref{fig:idefix_correlations}), hence they are both securely detected. The orbital eccentricity of both planets is consistent with zero. The posterior probability distributions over almost all the parameters, including those not shown on Figure~\ref{fig:idefix_correlations}, are approximately Gaussian and there are no significant correlations between parameters that might indicate a problem with the fit. The only exceptions are the impact parameters of the two planets and their eccentricity parameters $e \cos \omega$ and $e \sin \omega$, which do show some non-Gaussianity and mutual correlation, but this is entirely expected due to degeneracy between those parameter that cannot be fully resolved with these observations. The posterior distribution for the period of planet b (top left panel of Figure~\ref{fig:idefix_correlations}) is somewhat non-Gaussian, which could indicate a tension between the different data sets used in the analysis, in particular between the two \tess\ sectors as these  constrain the period of b more robustly than the RVs. We will return to this in Section~\ref{sec:dynamics} when discussing the system dynamics and the possibility of transit timing variations.

The fit includes a `jitter' term for each data type (photometry and RV), which is added in quadrature to the formal uncertainties. The photometric jitter term (285\,ppm) was found to be significantly smaller than the median photometric uncertainty (484\,ppm), indicating a good fit. The best-fit RV jitter was 2.9\,\ms, compared to a median formal RV uncertainty of 1.7\,m/s. This indicates either that the formal uncertainties were slightly underestimated, or that there is an additional source of uncorrelated or correlated signal, such as stellar phenomena (e.g. granulation), instrumental changes or additional planet signals, which are not accounted for by our model. This explains the comparatively large number of observations that were required to detect the RV signals of both planets to better than $4 \sigma$, despite the strong constraints on the orbital ephemerides provided by the transits.

As a final check, we computed the Lomb-Scargle periodogram of the RV residuals, which is shown as the purple line in Figure~\ref{fig:RV_pgram}. There are no significant peaks in the residuals, indicating no detection of additional planets in the system and or periodic signals correlated with our estimated stellar rotation period. 

\subsection{Planet masses and radii}

\begin{figure} 
    \centering
    \includegraphics[width=\linewidth]{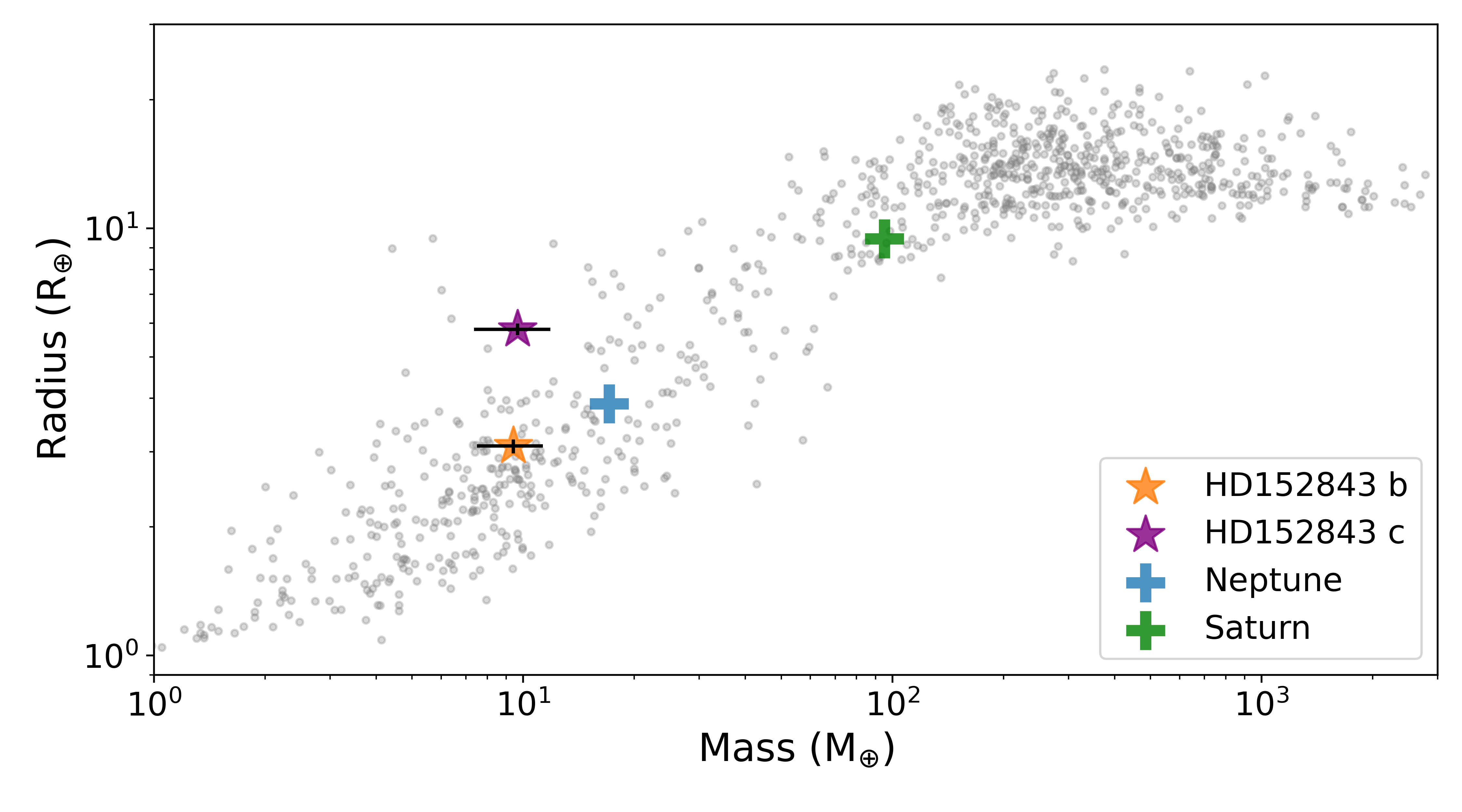}
    \caption{The plot show mass versus radius of HD152843 b and c compared with the sample of known planets with masses and radii known to better than 50\% precision, sourced from the TEPCAT database. For reference, Neptune and Saturn are shown as blue and green pluses, respectively.}
    \label{fig:mr_plot}
\end{figure}

We used the parameters of the transit and RV fit, reported in Table~\ref{tab:parstarget}, together with the stellar mass and radius from E21, to derive the bulk and orbital parameters of each planet. These are reported in Table~\ref{tab:derparstarget}. For planet b we find a mass of \mpb[]\,$M_{\oplus}$ and radius of \rpb[]\,$R_{\oplus}$, and for planet c we find a mass of \mpc[]\,$M_{\oplus}$ and a radius of \rpc[]\,$R_{\oplus}$. 

\begin{table}
\centering
  \caption{Derived system parameters. \label{tab:derparstarget}}  
  \begin{tabular}{lc}
  \hline
  Parameter & Value$^{(\mathrm{b})}$\\
  \hline
    \multicolumn{2}{l}{\textit{\textbf{Derived parameters \target b}}} \\
  \noalign{\smallskip}
    Planet mass ($M_{\oplus}$)    & \mpb[] \\
    & \\
    Planet radius ($R_{\oplus}$)    & \rpb[] \\
    & \\
    Planet density $\rho$ (${\rm g\,cm^{-3}}$)   & \denpb[] \\
    & \\
    Semi-major axis $a$ (AU)    & \ab[] \\
    & \\
    Eccentricity $e$   & \eb[] \\
    & \\
    Transit duration $\tau$ (hours)   & \ttotb[] \\
    & \\
    Orbit inclination $i$ (deg)    & \ib[] \\
    & \\
    Instellation $F_{\rm p}$ ($F_{\oplus}$)     & \insolationb[] \\
    \multicolumn{2}{l}{\textit{\textbf{Derived parameters \target c}}}
    \\
  \noalign{\smallskip}
    Planet mass ($M_{\oplus}$)    & \mpc[] \\
    & \\
    Planet radius ($R_{\oplus}$)    & \rpc[] \\
    & \\
    Planet density $\rho$ (${\rm g\,cm^{-3}}$)   & \denpb[] \\
    & \\
    Eccentricity $e$   & \ec[] \\
    & \\
    semi-major axis $a$ (AU)    & \ac[] \\
    & \\
    Transit duration $\tau$ (hours)  & \ttotc[] \\
    & \\
    Orbit inclination $i$ (deg)    & \ic[] \\
    & \\
    Instellation $F_{\rm p}$ ($F_{\oplus}$)     & \insolationc[] \\
    \hline
   \noalign{\smallskip}
  \end{tabular}
\end{table}

The masses and radii of the two planets are shown in Figure~\ref{fig:mr_plot}, where we compare them to published values for other transiting exoplanets from the TEPCAT catalogue \citep[][sourced 18/9/23]{tepcat}. Both planets have very similar masses, but very different radii, and hence densities. While Planet b is a fairly typical sub-Neptune, planet c, by contrast, belongs to the much rarer category of `super-puffs' -- extremely low density sub-Neptune mass planets. The other planets that


\section{Discussion}
\label{sec:discussion}

Having determined the period, masses and radii of both planets, we can now explore the implications of our results for the system's formation and evolution history, and the insights it can offer to the `super-puff' class of exoplanet. 

\subsection{Formation and Evolution}
\label{sec:MESAplanet}

The presence of two planets with such similar masses but different densities, both orbiting relatively close to their host star, is potentially puzzling - did the planets form with similar densities but subsequently evolve differently, or are these differences primordial? 

To explore the formation and evolution history of this intriguing system, we ran the {\sc MESA Planet} planetary evolution simulation \citep{Owen2020} to explore the range of possible initial states of planet c. This simulation adapts the Modules for Experiments in Stellar Astrophysics \citep[MESA][]{Paxton2011, Paxton2013, Paxton2015, Paxton2018, Paxton2019, Jermyn2023} stellar evolution code to include a planetary core of a specified density, and an external flux source to emulate the behaviour of the host star that evolves according to a theoretical model. For this simulation, we use a composite core of $\frac{2}{3}$ silicate and $\frac{1}{3}$ iron, and chose a MIST stellar model that most closely represents \target\ with a stellar mass of 1.16\,$M_{\odot}$ and metallicity of 0.25. 

\begin{figure} 
    \centering
    \includegraphics[width=\linewidth]{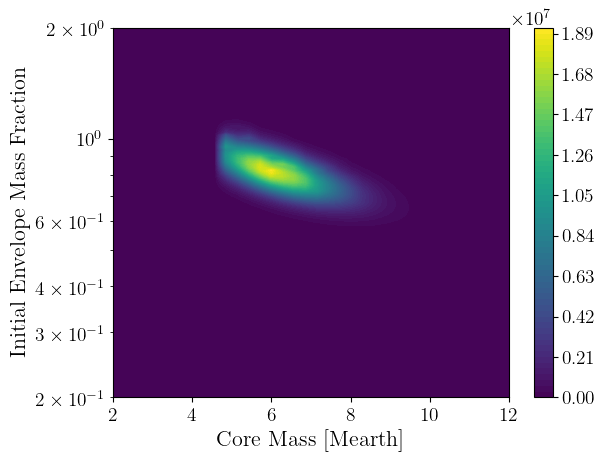}
    \includegraphics[width=\linewidth]{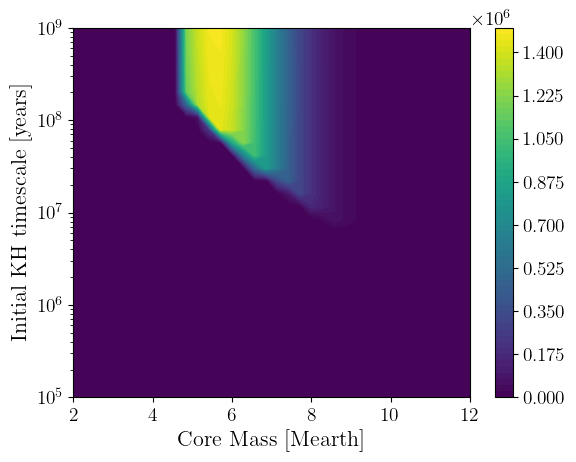}
    \includegraphics[width=\linewidth]{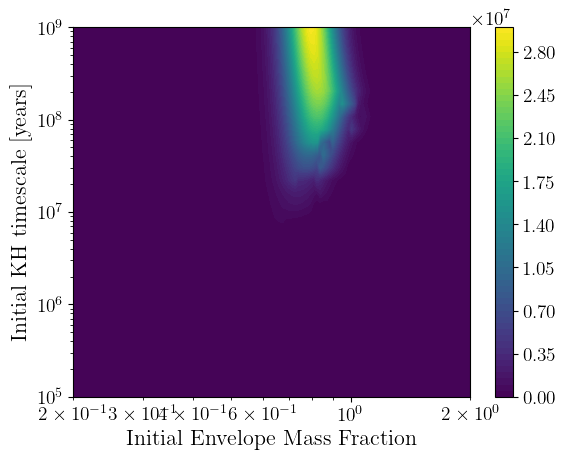}
    \caption{Results of the MESA planet modelling showing the likelihood contours for the three dimensional parameter space of core mass, initial envelope mass fraction and initial internal heating (parameterised as the Kelvin-Helmholtz cooling timescale). The lack of a clear, Gaussian-like maximum likelihood, and the tendency of the solution to push to higher and higher KH timescales beyond values that are physical indicate that the energy from initial formation and stellar irradiation alone is not enough to explain the current measured properties of \planetc.  }
    \label{fig:MESA_correlation}
\end{figure}

To determine the possible initial characteristics of planet c, simulations are run over a grid of varying core masses, envelope masses, and amounts of internal heating. Once the models are initialised and settled to a steady state, they are evolved starting at a stellar age of 3\,Myrs - the time at which it is thought that these types of planets have finished forming and the disc has cleared \citep{Hernandez2007, Mamajek2009} - to the current estimated age of $\sim 4$\,Gyrs. We then select the models that have final radii within 5 $\sigma$ of our solution and compute the likelihood over the grid of core masses, initial envelope mass (expressed as a fraction of the core mass) and internal heating (given in terms of the Kelvin-Helmholtz (KH) timescale - the time required for the planet to irradiate away the energy equivalent to its thermal energy), given our current estimates of the stellar age and planet mass, radius and equilibrium temperature. These likelihood contours are shown in figure \ref{fig:MESA_correlation}. The core mass versus initial envelope mass fraction contours (top panel) show a Gaussian-like maxima around a core mass of 6\,\MEarth and envelope mass fraction between 0.8 and 0.9 of the core mass. The other two likelihood contours do not show a clear maxima as the initial KH timescale is tending to higher and higher values, reaching the fixed upper-limit imposed as the highest physically meaningful value of initial kinetic energy for a forming planet. 
These results indicate that the initial heat from formation and stellar irradiation are not sufficient to explain the current properties of \planetc, and that an additional source of heating is required to explain the planet's low density. This is assuming that the relatively large radius for \planetc\, is not the result of high-altitude haze, as this is not in the scope of these models. 

\subsection{System Dynamics}
\label{sec:dynamics}

The ratio of the orbital periods of planets b and c ($P_c/P_b = 1.678220 \pm 1.4e-5 $) places these planets within 7\% of the 5:3 mean motion resonance (MMR). This is a second order resonance, meaning it is not as energetically favourable or as common \citep{Lissauer2011} as a first order 2:1 or 3:2 resonance. \cite{Pan2020} demonstrate that these higher order resonances can occur in systems where two inner planets dynamically interact with an outer giant planet. Further, they showed that a near 5:3 resonance in particular is made possible if the giant planet is on a eccentric orbit. This suggests that there may be, or may have been, a third planet in the \target\ system, though no such third planet is seen in the RV or photometric data presented here.

This near MMR may give rise to observable Transit Timing Variations (TTVs). The gravitational interactions of n-body systems cause them to deviate from Keplerian orbits \citep{Agol2005, Holman2005}, which can be seen in photometric time series data. This usually takes the form of evolving, sinusoidal, periodic offsets in a planet's transit times compared to an average linear ephemeris. These offsets can occur on the order of minutes to hours.

\begin{figure} 
    \centering
    \includegraphics[width=.5\textwidth]{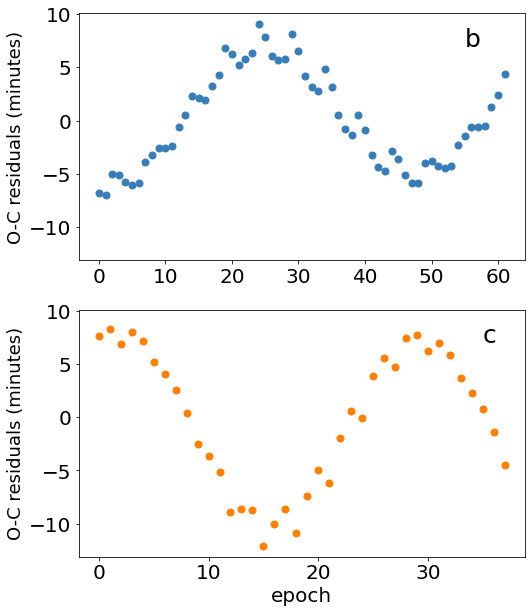}   
    \caption{The expected Observed minus Calculated (O-C) residuals for planet b (top) and c (bottom) for planetary parameters obtained in this study, which show periodic TTVs with detectable amplitudes. Significant scatter or "chopping" on the order of minutes can be seen in the timings of both planets.}
    \label{fig:TTVs_sim}
\end{figure}

\begin{figure} 
    \centering
    \includegraphics[width=.5\textwidth]{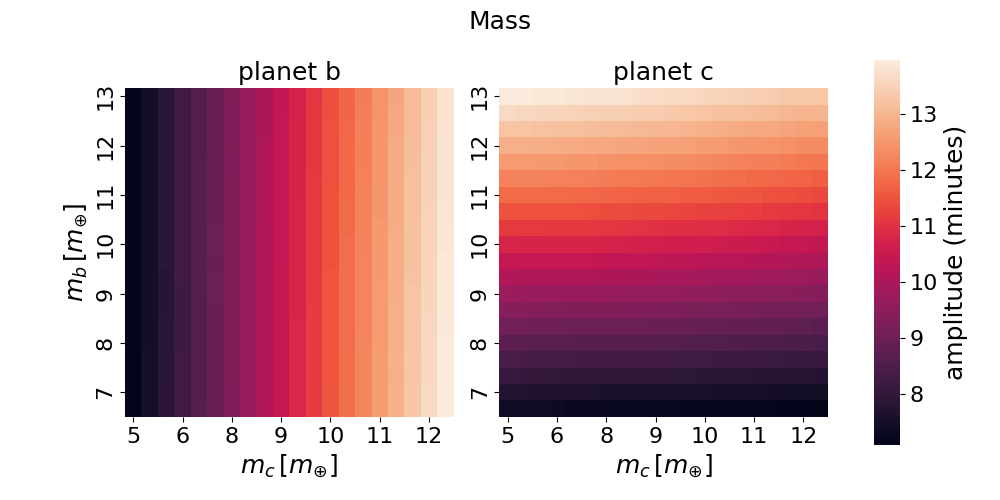} 
    \includegraphics[width=.5\textwidth]{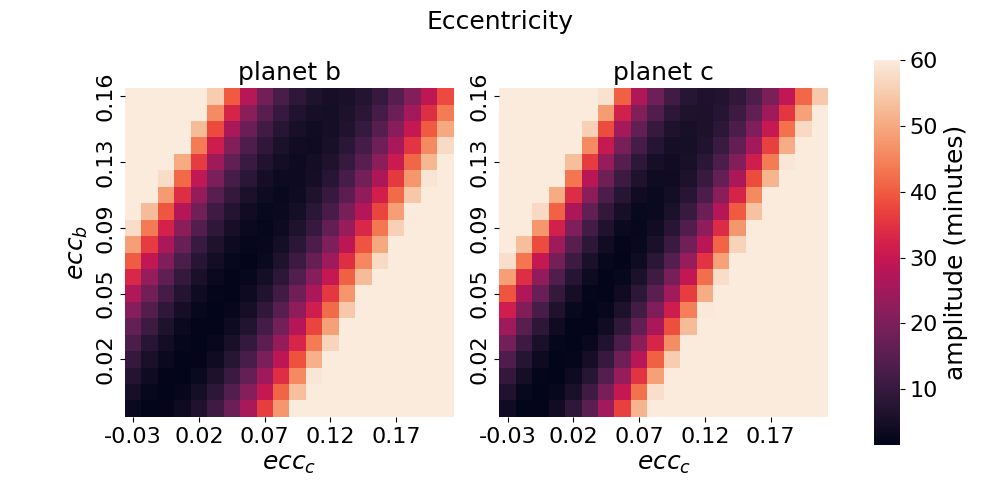}    
    \caption{Heatmaps show expected amplitudes of TTVs (O-C offsets) for a range of parameters across our uncertainty distribution. The effect of varying the mass of each planet to two sigma plus or minus is shown in the top panel, and the effect of varying eccentricities is shown in the bottom panel. For most combinations, TTVs remain detectable at amplitudes of 10 minutes or more. In each case, the parameter in question is varied while all other parameters are fixed at their median value.}
    \label{fig:TTV_heatmaps}
\end{figure}

We simulated the expected TTV signal given the planetary system parameters obtained above using the TTVFast package \citep{Deck2014b}. Because of the proximity of the planets' periods to a MMR, a periodic signal is expected to be observed with an amplitude of 15-20 minutes, as seen in Figure \ref{fig:TTVs_sim}, which shows the simulated residuals to a linear ephemeris of each planet over a superperiod. While uncertainties exist on each of the input parameters, Figure \ref{fig:TTV_heatmaps} shows a range of TTV amplitudes across a two sigma distribution of input parameters, with the most significant influence coming from planetary masses and eccentricities. As expected, the signal is most sensitive to eccentricity, with higher eccentricities leading to larger TTVs. However, a wide range of masses and eccentricities in the two planets should still allow for TTVs to be detectable, given the precision of individual transit times for each planet on the order of 4 minutes and 2 minutes for planets b and c respectively.

The superperiod, $P_{TTV}$, of two interacting planets in j:j-i orbital resonance is given by 
\begin{equation}
    P_{TTV} = {\left|\frac{j-i}{P_{inner}} - \frac{j}{P_{outer}} \right|}^{-1}
\end{equation}
For \target \, this gives rise to a superperiod of approximately 560 days, which for planet b is on the order of 50 epochs. While the spread in transit times covers this superperiod, several more intermediate transits would be necessary to confirm the non-linearity of residuals.

\begin{figure} 
    \centering
    \includegraphics[width=.5\textwidth]{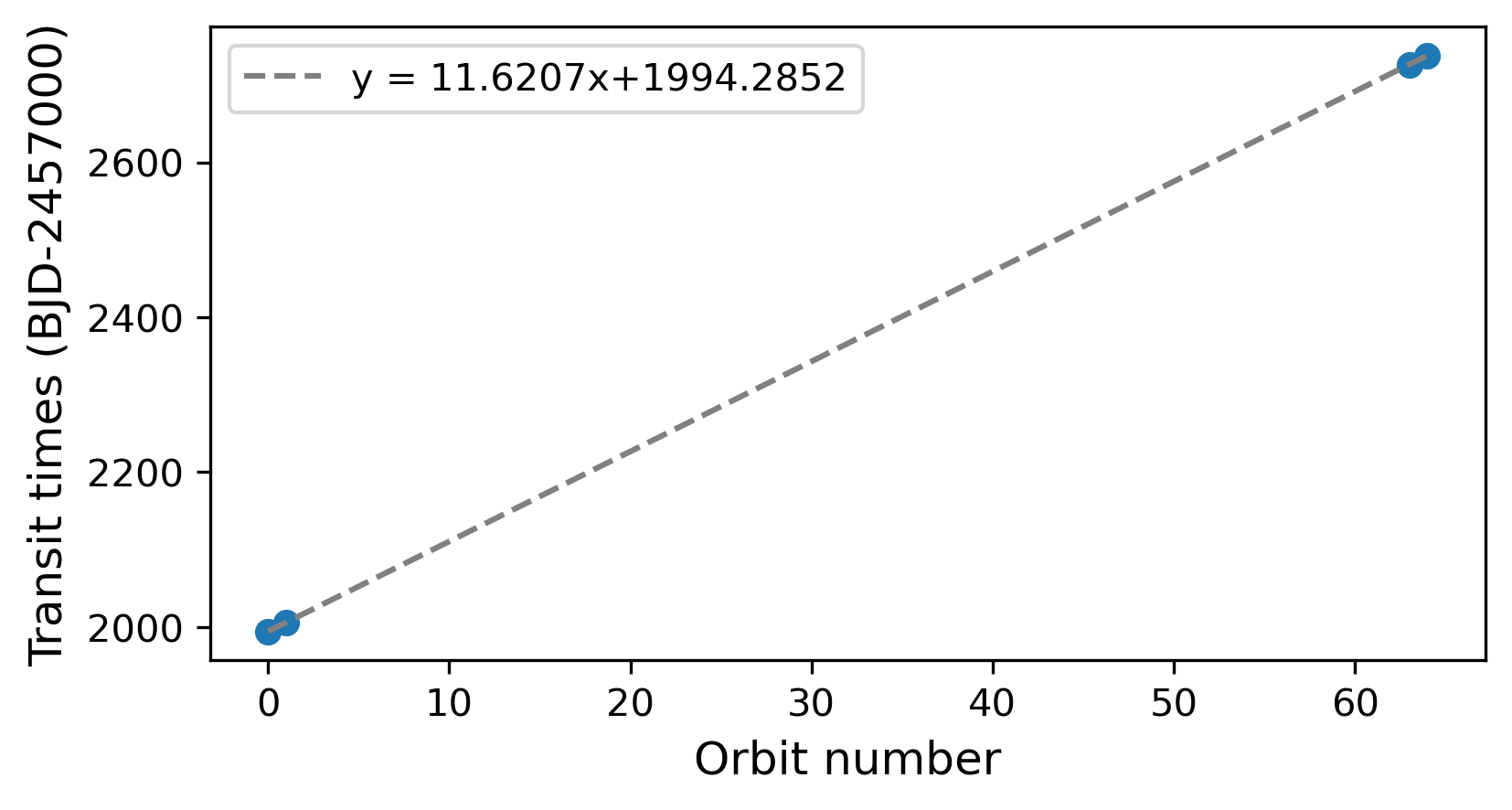} 
    \includegraphics[width=.5\textwidth]{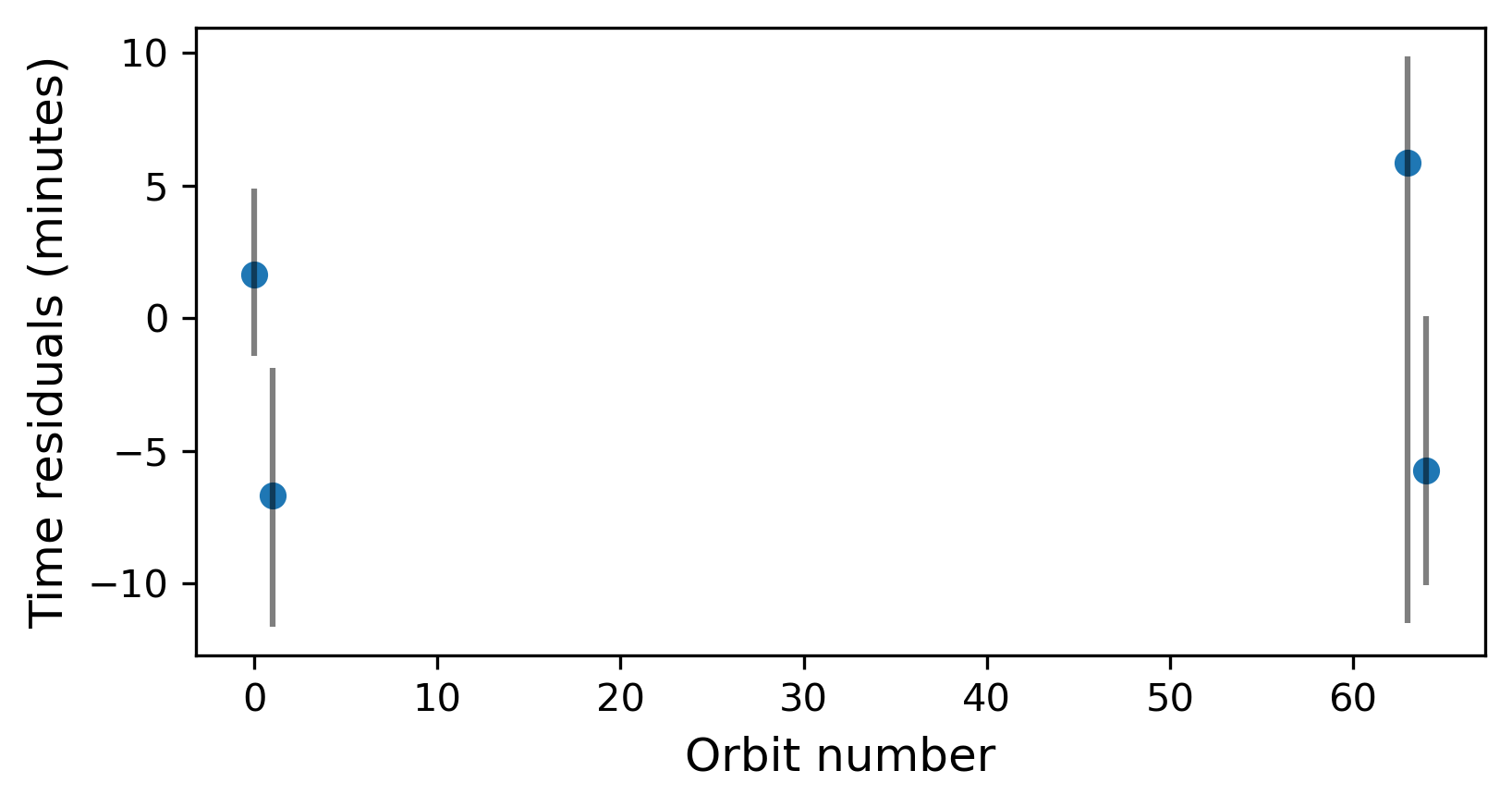}    
    \caption{Top panel: Transit times of \planetb in \tess\ data as a function of transit number. Bottom panel: residual in transit time compared to a linear ephemeris (perfectly Keplarian orbit). These data show no obvious signs of transit timing variations, however the scatter in the points in  large uncertainties do not rule them out. }
    \label{fig:TTVs}
\end{figure}

We checked for any tentative evidence of TTVs in the 4 transits observed so far for planet b, as follows. We fit each transit individually, keeping all the parameters fixed to the values given in Table~\ref{tab:parstarget}, and varying only the time of transit centre ($T_0$). These values are plotted in the top panel of Figure~\ref{fig:TTVs}, along with the linear ephemeris determined by out joint fit in Section \ref{sec:pyaneti}. The transit time residuals from this linear ephemeris are shown in the bottom panel of Figure~\ref{fig:TTVs}. There are no TTVs obviously present in this data, however, given the uncertainties of each time point, TTVs are not ruled out. Further, the scatter currently observed in mid-point times of the four transits of planet b is consistent with chopping effects, which are the short timescale offsets associated with the planet's synodic period evolution \citep{Deck2014}. More transit observations are needed to confirm the presence of any TTVs and enable us to better constrain planetary masses and orbital properties. 

\subsection{Prospects for atmospheric characterisation}

\begin{figure*} 
    \centering
    \includegraphics[width=0.48\linewidth]{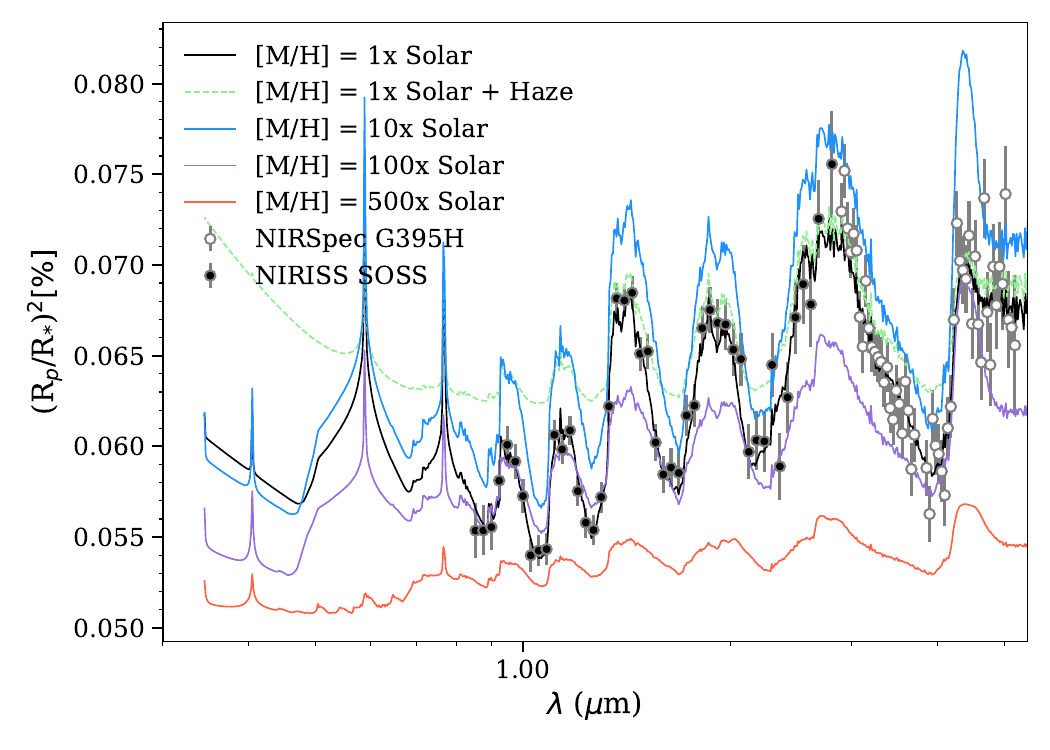} \hfill
    \includegraphics[width=0.48\linewidth]{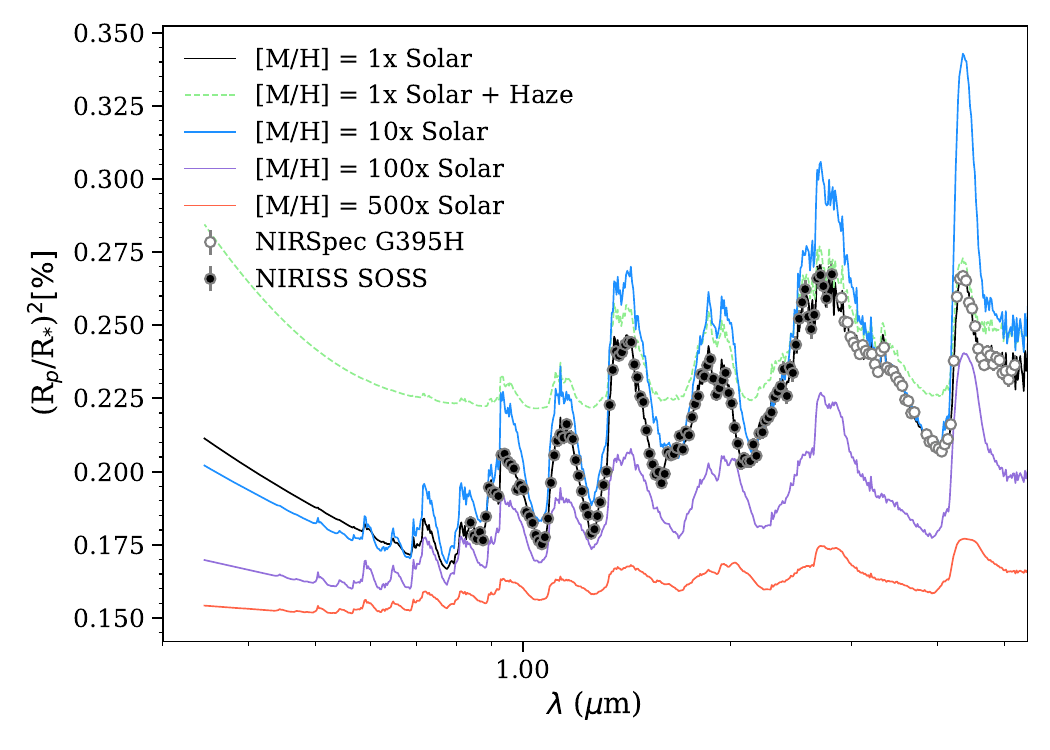}    
    \caption{Simulated transmission spectra for planets b (left) and c (right) for different assumptions regarding metallicity and the presence of a haze. The points with error bars show single-transit observations with JWST observations simulated using Pandexo, demonstrating that molecular features would be detected at high confidence irrespective of those assumptions, particularly using the NIRSpec/G395H grism.}
    \label{fig:JWST}
\end{figure*}

The combination of a bright host star and two Neptune-mass transiting planets makes the \target\ system an ideal target for atmospheric studies and comparative planetology.
\planetc\ in particular is an outstanding target for transmission spectroscopy owing to its extremely low density, and has very high transmission spectrum metric \citep[TSM, as defined in][]{Kempton2018} of 304, and \planetb\ has a lower TSM of 53. This places planet c well above the threshold TSM of 90 for 'high-quality atmospheric characterisation' as defined by \cite{Kempton2018}. By measuring the transmission spectra of both planets over an appropriate wavelength range, one could constrain the metalicities and C/O ratios of their atmospheres, and, more importantly, allow us to determine the reason for the large radius of planet c by indicating the presence or lack thereof of high-altitude haze. 

To illustrate the observable potential of planets b and c with JWST, we computed a suite of forward models using the open-source code CHIMERA \citep{Line2013} and compared these to synthetic observations of each planet that were generated using PandExo \citep{Batalha2019}. We use PandExo to model 1 transit of each planet using NIRISS/SOSS and NIRSpec/G395H, thus spanning multiple H$_2$O bands and important carbon-bearing molecules such as CO$_2$ and CH$_4$. We use the newly derived system parameters presented in Table \ref{tab:derparstarget}. We assume an isothermal temperature structure (values taken from Table 1 in \citet{Eisner2021}) and model the atmosphere without clouds. We model the atmospheres to have a solar C/O ratio and metalicities of 1$\times$, 10$\times$, 100$\times$, 300$\times$ and 500$\times$ solar respectively. We generated an additional model with haze for our 1$\times$ solar metalicity model, to highlight how haze would impact the transmission spectrum. We used the chemical grid developed in \citep{Kreidberg2015}. These models can be seen in Figure \ref{fig:JWST}, we over-plot our synthetic observations for the 1$\times$ solar metalicity model. The black data points are the synthetic NIRISS/SOSS observations and the white points are the synthetic NIRSpec/G395H data points. It is important to use both of these observing modes as they span key parts of the transmission spectrum. NIRISS/SOSS probes the haziness of the atmosphere, high altitude hazes would mute the H$_2$O bands in this region. NIRSpec/G395H spans a H$_2$O band, CO$_2$ band and CH$_4$ band, thus possible to measure the C/O of the atmosphere.

\section{Summary and Conclusions}
\label{sec:sum_concl}
This work presents an intensive campaign of radial velocity observations with HARPS-N to measure the masses of the two transiting planets discovered by citizens around \target.  Additional \tess\ data that has become available since the initial publication of this system, has allows us to refine the radii estimates and orbital periods of both planets. All derived planet properties are summaried in Table \ref{tab:derparstarget}. 

These mass measurements reveal a fascinating planetary system: a low density, super-puff outer planet in a likely second-order resonant orbit with a typical Neptune-size inner planet. Such a system allows us to explore scenarios of formation of super-puff planets, and the properties of planetary systems with planets near higher order mean motion resonances. Modelling the formation and evolution of planet c shows that, if the large radius is not due to high-altitude haze, it is unlikely that internal heat from formation and host star irradiation alone are enough to explain its current size. In such a case, an extra source of heat, such as that caused by dynamical interactions between planet b and c, is needed to explain planet c's inflated radius. Future observations with JWST and other space-based instruments will allow us to explore the nature of the atmospheres for both targets and will reveal if planet c's large radius is caused by high-altitude hazes or if something more complex is occurring with this system. Future TTV observations and dynamical modelling of this system will give insight to the possible formation and migration scenarios of this system, and the possibility of a distant third planet in the system not seen in the radial velocity or photometric data to date. 


\section*{Acknowledgements}

BN would like to acknowledge support from STFC Consolidated Grant ST/S000488/1 (PI Balbus), and the University of Southern Queensland SAGE program. BN would also like to thank Dr. Su Wang for her insights on planetary system dynamics. ACC acknowledges support from STFC consolidated grant number ST/V000861/1, and UKSA grant number ST/R003203/1.

\section*{Data Availability}

The \tess\ data are publicly available via the Barbara A. Mikulski Archive for Space Telescopes\footnote{\url{https://mast.stsci.edu/}} (MAST). The HARPS-N DRS data published here are publicly available via the Data \& Analysis Centre for Exoplanets\footnote{\url{https://dace.unige.ch/}} (DACE), and our full table of the DRS and YARARA RVs and data products are available as supplementary material and on Vizier CDS.  All other data and analysis products used in this paper are available on request. 
 



\bibliographystyle{mnras}
\bibliography{HD152843Masses} 

\begin{thebibliography}{}
\makeatletter
\relax
\def\mn@urlcharsother{\let\do\@makeother \do\$\do\&\do\#\do\^\do\_\do\%\do\~}
\def\mn@doi{\begingroup\mn@urlcharsother \@ifnextchar [ {\mn@doi@}
  {\mn@doi@[]}}
\def\mn@doi@[#1]#2{\def\@tempa{#1}\ifx\@tempa\@empty \href
  {http://dx.doi.org/#2} {doi:#2}\else \href {http://dx.doi.org/#2} {#1}\fi
  \endgroup}
\def\mn@eprint#1#2{\mn@eprint@#1:#2::\@nil}
\def\mn@eprint@arXiv#1{\href {http://arxiv.org/abs/#1} {{\tt arXiv:#1}}}
\def\mn@eprint@dblp#1{\href {http://dblp.uni-trier.de/rec/bibtex/#1.xml}
  {dblp:#1}}
\def\mn@eprint@#1:#2:#3:#4\@nil{\def\@tempa {#1}\def\@tempb {#2}\def\@tempc
  {#3}\ifx \@tempc \@empty \let \@tempc \@tempb \let \@tempb \@tempa \fi \ifx
  \@tempb \@empty \def\@tempb {arXiv}\fi \@ifundefined
  {mn@eprint@\@tempb}{\@tempb:\@tempc}{\expandafter \expandafter \csname
  mn@eprint@\@tempb\endcsname \expandafter{\@tempc}}}

\bibitem[\protect\citeauthoryear{{Agol}, {Steffen}, {Sari}  \&
  {Clarkson}}{{Agol} et~al.}{2005}]{Agol2005}
{Agol} E.,  {Steffen} J.,  {Sari} R.,   {Clarkson} W.,  2005, \mn@doi [\mnras]
  {10.1111/j.1365-2966.2005.08922.x}, \href
  {https://ui.adsabs.harvard.edu/abs/2005MNRAS.359..567A} {359, 567}

\bibitem[\protect\citeauthoryear{{Baranne} et~al.,}{{Baranne}
  et~al.}{1996}]{Baranne1996}
{Baranne} A.,  et~al., 1996, \aaps, \href
  {https://ui.adsabs.harvard.edu/abs/1996A&AS..119..373B} {119, 373}

\bibitem[\protect\citeauthoryear{{Barrag{\'a}n}, {Gandolfi}  \&
  {Antoniciello}}{{Barrag{\'a}n} et~al.}{2019}]{pyaneti1}
{Barrag{\'a}n} O.,  {Gandolfi} D.,   {Antoniciello} G.,  2019, \mn@doi [\mnras]
  {10.1093/mnras/sty2472}, \href
  {https://ui.adsabs.harvard.edu/abs/2019MNRAS.482.1017B} {482, 1017}

\bibitem[\protect\citeauthoryear{{Barrag{\'a}n}, {Aigrain}, {Rajpaul}  \&
  {Zicher}}{{Barrag{\'a}n} et~al.}{2022}]{pyaneti2}
{Barrag{\'a}n} O.,  {Aigrain} S.,  {Rajpaul} V.~M.,   {Zicher} N.,  2022,
  \mn@doi [\mnras] {10.1093/mnras/stab2889}, \href
  {https://ui.adsabs.harvard.edu/abs/2022MNRAS.509..866B} {509, 866}

\bibitem[\protect\citeauthoryear{{Batalha} et~al.,}{{Batalha}
  et~al.}{2019}]{Batalha2019}
{Batalha} N.~E.,  et~al., 2019, {PandExo: Instrument simulations for exoplanet
  observation planning}, Astrophysics Source Code Library, record ascl:1906.016
  (\mn@eprint {ascl} {1906.016})

\bibitem[\protect\citeauthoryear{{Bourrier} et~al.,}{{Bourrier}
  et~al.}{2021}]{Bourrier2021}
{Bourrier} V.,  et~al., 2021, \mn@doi [\aap] {10.1051/0004-6361/202141527},
  \href {https://ui.adsabs.harvard.edu/abs/2021A&A...654A.152B} {654, A152}

\bibitem[\protect\citeauthoryear{{Cosentino} et~al.,}{{Cosentino}
  et~al.}{2012}]{Cosentino2012}
{Cosentino} R.,  et~al., 2012, in {McLean} I.~S.,  {Ramsay} S.~K.,   {Takami}
  H.,  eds,  Society of Photo-Optical Instrumentation Engineers (SPIE)
  Conference Series Vol. 8446, Ground-based and Airborne Instrumentation for
  Astronomy IV. p. 84461V, \mn@doi{10.1117/12.925738}

\bibitem[\protect\citeauthoryear{{Cosentino} et~al.,}{{Cosentino}
  et~al.}{2014}]{Cosentino2014}
{Cosentino} R.,  et~al., 2014, in {Ramsay} S.~K.,  {McLean} I.~S.,   {Takami}
  H.,  eds,  Society of Photo-Optical Instrumentation Engineers (SPIE)
  Conference Series Vol. 9147, Ground-based and Airborne Instrumentation for
  Astronomy V. p. 91478C, \mn@doi{10.1117/12.2055813}

\bibitem[\protect\citeauthoryear{{Cretignier}, {Dumusque}, {Allart}, {Pepe}  \&
  {Lovis}}{{Cretignier} et~al.}{2020a}]{Cretignier2020a}
{Cretignier} M.,  {Dumusque} X.,  {Allart} R.,  {Pepe} F.,   {Lovis} C.,
  2020a, \mn@doi [\aap] {10.1051/0004-6361/201936548}, \href
  {https://ui.adsabs.harvard.edu/abs/2020A&A...633A..76C} {633, A76}

\bibitem[\protect\citeauthoryear{{Cretignier}, {Francfort}, {Dumusque},
  {Allart}  \& {Pepe}}{{Cretignier} et~al.}{2020b}]{Cretignier2020b}
{Cretignier} M.,  {Francfort} J.,  {Dumusque} X.,  {Allart} R.,   {Pepe} F.,
  2020b, \mn@doi [\aap] {10.1051/0004-6361/202037722}, \href
  {https://ui.adsabs.harvard.edu/abs/2020A&A...640A..42C} {640, A42}

\bibitem[\protect\citeauthoryear{{Cretignier}, {Dumusque}, {Hara}  \&
  {Pepe}}{{Cretignier} et~al.}{2021}]{Cretignier2021}
{Cretignier} M.,  {Dumusque} X.,  {Hara} N.~C.,   {Pepe} F.,  2021, \mn@doi
  [\aap] {10.1051/0004-6361/202140986}, \href
  {https://ui.adsabs.harvard.edu/abs/2021A&A...653A..43C} {653, A43}

\bibitem[\protect\citeauthoryear{{Deck} \& {Agol}}{{Deck} \&
  {Agol}}{2014}]{Deck2014}
{Deck} K.,  {Agol} E.,  2014, \mn@doi [\apj] {10.1088/0004-637X/802/2/116}

\bibitem[\protect\citeauthoryear{{Deck}, {Agol}, {Holman}  \&
  {Nesvorn{\'y}}}{{Deck} et~al.}{2014}]{Deck2014b}
{Deck} K.~M.,  {Agol} E.,  {Holman} M.~J.,   {Nesvorn{\'y}} D.,  2014, \mn@doi
  [\apj] {10.1088/0004-637X/787/2/132}, \href
  {https://ui.adsabs.harvard.edu/abs/2014ApJ...787..132D} {787, 132}

\bibitem[\protect\citeauthoryear{{Dumusque} et~al.,}{{Dumusque}
  et~al.}{2021}]{Dumusque2021}
{Dumusque} X.,  et~al., 2021, \mn@doi [\aap] {10.1051/0004-6361/202039350},
  \href {https://ui.adsabs.harvard.edu/abs/2021A&A...648A.103D} {648, A103}

\bibitem[\protect\citeauthoryear{{Eisner} et~al.,}{{Eisner}
  et~al.}{2021a}]{eisner2020method}
{Eisner} N.~L.,  et~al., 2021a, \mn@doi [\mnras] {10.1093/mnras/staa3739},
  \href {https://ui.adsabs.harvard.edu/abs/2021MNRAS.501.4669E} {501, 4669}

\bibitem[\protect\citeauthoryear{{Eisner} et~al.,}{{Eisner}
  et~al.}{2021b}]{Eisner2021}
{Eisner} N.~L.,  et~al., 2021b, \mn@doi [\mnras] {10.1093/mnras/stab1253},
  \href {https://ui.adsabs.harvard.edu/abs/2021MNRAS.505.1827E} {505, 1827}

\bibitem[\protect\citeauthoryear{{Fausnaugh} et~al.,}{{Fausnaugh}
  et~al.}{2022}]{Sector52Release}
{Fausnaugh} M.~M.,  et~al., 2022, Technical report, TESS Data Release Notes:
  Sector 52, DR76, \url
  {https://tasoc.dk/docs/release_notes/tess_sector_52_drn76_v01.pdf}.
NASA, \url {https://tasoc.dk/docs/release_notes/tess_sector_52_drn76_v01.pdf}

\bibitem[\protect\citeauthoryear{{Gao} \& {Zhang}}{{Gao} \&
  {Zhang}}{2020}]{GaoZhang2020}
{Gao} P.,  {Zhang} X.,  2020, \mn@doi [\apj] {10.3847/1538-4357/ab6a9b}, \href
  {https://ui.adsabs.harvard.edu/abs/2020ApJ...890...93G} {890, 93}

\bibitem[\protect\citeauthoryear{{Gelman} \& {Rubin}}{{Gelman} \&
  {Rubin}}{1992}]{Gelman1992}
{Gelman} A.,  {Rubin} D.~B.,  1992, \mn@doi [Statistical Science]
  {10.1214/ss/1177011136}, \href
  {https://ui.adsabs.harvard.edu/abs/1992StaSc...7..457G} {7, 457}

\bibitem[\protect\citeauthoryear{Gelman, Carlin, Stern  \& Rubin}{Gelman
  et~al.}{2003}]{Gelman2003}
Gelman A.,  Carlin J.,  Stern H.,   Rubin D.,  2003, Bayesian Data Analysis,
  Second Edition.
Chapman \& Hall/CRC Texts in Statistical Science, Taylor \& Francis, \url
  {https://books.google.com.au/books?id=TNYhnkXQSjAC}

\bibitem[\protect\citeauthoryear{{Hara}, {Bou{\'e}}, {Laskar}  \&
  {Correia}}{{Hara} et~al.}{2017}]{Hara(2017)}
{Hara} N.~C.,  {Bou{\'e}} G.,  {Laskar} J.,   {Correia} A.~C.~M.,  2017,
  \mn@doi [\mnras] {10.1093/mnras/stw2261}, \href
  {https://ui.adsabs.harvard.edu/abs/2017MNRAS.464.1220H} {464, 1220}

\bibitem[\protect\citeauthoryear{{Helled}}{{Helled}}{2023}]{Helled2023}
{Helled} R.,  2023, \mn@doi [\aap] {10.1051/0004-6361/202346850}, \href
  {https://ui.adsabs.harvard.edu/abs/2023A&A...675L...8H} {675, L8}

\bibitem[\protect\citeauthoryear{{Hern{\'a}ndez} et~al.,}{{Hern{\'a}ndez}
  et~al.}{2007}]{Hernandez2007}
{Hern{\'a}ndez} J.,  et~al., 2007, \mn@doi [\apj] {10.1086/522882}, \href
  {https://ui.adsabs.harvard.edu/abs/2007ApJ...671.1784H} {671, 1784}

\bibitem[\protect\citeauthoryear{Holman \& Murray}{Holman \&
  Murray}{2005}]{Holman2005}
Holman M.~J.,  Murray N.~W.,  2005, \mn@doi [Science]
  {10.1126/science.1107822}, 307, 1288

\bibitem[\protect\citeauthoryear{{Jenkins} et~al.,}{{Jenkins}
  et~al.}{2016}]{Jenkins2016}
{Jenkins} J.~M.,  et~al., 2016, in \procspie. p. 99133E,
  \mn@doi{10.1117/12.2233418}

\bibitem[\protect\citeauthoryear{{Jermyn} et~al.,}{{Jermyn}
  et~al.}{2023}]{Jermyn2023}
{Jermyn} A.~S.,  et~al., 2023, \mn@doi [\apjs] {10.3847/1538-4365/acae8d},
  \href {https://ui.adsabs.harvard.edu/abs/2023ApJS..265...15J} {265, 15}

\bibitem[\protect\citeauthoryear{{Kempton} et~al.,}{{Kempton}
  et~al.}{2018}]{Kempton2018}
{Kempton} E. M.~R.,  et~al., 2018, \mn@doi [\pasp] {10.1088/1538-3873/aadf6f},
  \href {https://ui.adsabs.harvard.edu/abs/2018PASP..130k4401K} {130, 114401}

\bibitem[\protect\citeauthoryear{{Kipping}, {Spiegel}  \& {Sasselov}}{{Kipping}
  et~al.}{2013}]{Kipping2013}
{Kipping} D.~M.,  {Spiegel} D.~S.,   {Sasselov} D.~D.,  2013, \mn@doi [\mnras]
  {10.1093/mnras/stt1050}, \href
  {https://ui.adsabs.harvard.edu/abs/2013MNRAS.434.1883K} {434, 1883}

\bibitem[\protect\citeauthoryear{{Kreidberg} et~al.,}{{Kreidberg}
  et~al.}{2015}]{Kreidberg2015}
{Kreidberg} L.,  et~al., 2015, \mn@doi [\apj] {10.1088/0004-637X/814/1/66},
  \href {https://ui.adsabs.harvard.edu/abs/2015ApJ...814...66K} {814, 66}

\bibitem[\protect\citeauthoryear{{Lammer} et~al.,}{{Lammer}
  et~al.}{2016}]{Lammer2016}
{Lammer} H.,  et~al., 2016, \mn@doi [\mnras] {10.1093/mnrasl/slw095}, \href
  {https://ui.adsabs.harvard.edu/abs/2016MNRAS.461L..62L} {461, L62}

\bibitem[\protect\citeauthoryear{{Lightkurve Collaboration}
  et~al.,}{{Lightkurve Collaboration} et~al.}{2018}]{Lightkurve2018}
{Lightkurve Collaboration} et~al., 2018, {Lightkurve: Kepler and TESS time
  series analysis in Python}, Astrophysics Source Code Library (\mn@eprint
  {ascl} {1812.013})

\bibitem[\protect\citeauthoryear{{Line} et~al.,}{{Line}
  et~al.}{2013}]{Line2013}
{Line} M.~R.,  et~al., 2013, \mn@doi [\apj] {10.1088/0004-637X/775/2/137},
  \href {https://ui.adsabs.harvard.edu/abs/2013ApJ...775..137L} {775, 137}

\bibitem[\protect\citeauthoryear{{Lissauer} et~al.,}{{Lissauer}
  et~al.}{2011}]{Lissauer2011}
{Lissauer} J.~J.,  et~al., 2011, \mn@doi [\apjs] {10.1088/0067-0049/197/1/8},
  \href {https://ui.adsabs.harvard.edu/abs/2011ApJS..197....8L} {197, 8}

\bibitem[\protect\citeauthoryear{{Mamajek}}{{Mamajek}}{2009}]{Mamajek2009}
{Mamajek} E.~E.,  2009, in {Usuda} T.,  {Tamura} M.,   {Ishii} M.,  eds,
  American Institute of Physics Conference Series Vol. 1158, Exoplanets and
  Disks: Their Formation and Diversity. pp 3--10 (\mn@eprint {arXiv}
  {0906.5011}), \mn@doi{10.1063/1.3215910}

\bibitem[\protect\citeauthoryear{{Mandel} \& {Agol}}{{Mandel} \&
  {Agol}}{2002}]{Mandel2002}
{Mandel} K.,  {Agol} E.,  2002, \mn@doi [\apjl] {10.1086/345520}, \href
  {https://ui.adsabs.harvard.edu/abs/2002ApJ...580L.171M} {580, L171}

\bibitem[\protect\citeauthoryear{{Millholland}, {Petigura}  \&
  {Batygin}}{{Millholland} et~al.}{2020}]{Millholland2020}
{Millholland} S.,  {Petigura} E.,   {Batygin} K.,  2020, \mn@doi [\apj]
  {10.3847/1538-4357/ab959c}, \href
  {https://ui.adsabs.harvard.edu/abs/2020ApJ...897....7M} {897, 7}

\bibitem[\protect\citeauthoryear{{Owen}}{{Owen}}{2020}]{Owen2020}
{Owen} J.~E.,  2020, \mn@doi [\mnras] {10.1093/mnras/staa2784}, \href
  {https://ui.adsabs.harvard.edu/abs/2020MNRAS.498.5030O} {498, 5030}

\bibitem[\protect\citeauthoryear{{Pan}, {Wang}  \& {Ji}}{{Pan}
  et~al.}{2020}]{Pan2020}
{Pan} M.,  {Wang} S.,   {Ji} J.,  2020, \mn@doi [\mnras]
  {10.1093/mnras/staa1884}, \href
  {https://ui.adsabs.harvard.edu/abs/2020MNRAS.496.4688P} {496, 4688}

\bibitem[\protect\citeauthoryear{{Paxton}, {Bildsten}, {Dotter}, {Herwig},
  {Lesaffre}  \& {Timmes}}{{Paxton} et~al.}{2011}]{Paxton2011}
{Paxton} B.,  {Bildsten} L.,  {Dotter} A.,  {Herwig} F.,  {Lesaffre} P.,
  {Timmes} F.,  2011, \mn@doi [\apjs] {10.1088/0067-0049/192/1/3}, \href
  {https://ui.adsabs.harvard.edu/abs/2011ApJS..192....3P} {192, 3}

\bibitem[\protect\citeauthoryear{{Paxton} et~al.,}{{Paxton}
  et~al.}{2013}]{Paxton2013}
{Paxton} B.,  et~al., 2013, \mn@doi [\apjs] {10.1088/0067-0049/208/1/4}, \href
  {https://ui.adsabs.harvard.edu/abs/2013ApJS..208....4P} {208, 4}

\bibitem[\protect\citeauthoryear{{Paxton} et~al.,}{{Paxton}
  et~al.}{2015}]{Paxton2015}
{Paxton} B.,  et~al., 2015, \mn@doi [\apjs] {10.1088/0067-0049/220/1/15}, \href
  {https://ui.adsabs.harvard.edu/abs/2015ApJS..220...15P} {220, 15}

\bibitem[\protect\citeauthoryear{{Paxton} et~al.,}{{Paxton}
  et~al.}{2018}]{Paxton2018}
{Paxton} B.,  et~al., 2018, \mn@doi [\apjs] {10.3847/1538-4365/aaa5a8}, \href
  {https://ui.adsabs.harvard.edu/abs/2018ApJS..234...34P} {234, 34}

\bibitem[\protect\citeauthoryear{{Paxton} et~al.,}{{Paxton}
  et~al.}{2019}]{Paxton2019}
{Paxton} B.,  et~al., 2019, \mn@doi [\apjs] {10.3847/1538-4365/ab2241}, \href
  {https://ui.adsabs.harvard.edu/abs/2019ApJS..243...10P} {243, 10}

\bibitem[\protect\citeauthoryear{{Ricker} et~al.,}{{Ricker}
  et~al.}{2015}]{Ricker15}
{Ricker} G.~R.,  et~al., 2015, \mn@doi [Journal of Astronomical Telescopes,
  Instruments, and Systems] {10.1117/1.JATIS.1.1.014003}, \href
  {https://ui.adsabs.harvard.edu/abs/2015JATIS...1a4003R} {1, 014003}

\bibitem[\protect\citeauthoryear{{Southworth}}{{Southworth}}{2011}]{tepcat}
{Southworth} J.,  2011, \mn@doi [\mnras] {10.1111/j.1365-2966.2011.19399.x},
  \href {https://ui.adsabs.harvard.edu/abs/2011MNRAS.417.2166S} {417, 2166}

\makeatother
\end{thebibliography}

\bsp	
\label{lastpage}
\end{document}